\documentclass[sigplan,twocolumn]{acmart}

\usepackage{cleveref}
\usepackage{pdfpages}
\usepackage{subcaption}
\usepackage{float}
\usepackage{hyperref}
\usepackage{multirow}

\setcopyright{none}
\copyrightyear{2025}
\acmYear{2025}
\acmDOI{XXXXXXX.XXXXXXX}
\acmConference[Conference acronym 'XX]{Make sure to enter the correct
  conference title from your rights confirmation email}{June 03--05,
  2018}{Woodstock, NY}
\acmISBN{978-1-4503-XXXX-X/2018/06}

\newcommand{\microscenery}{\anon[\emph{redacted for anonymous review}]{microscenery\{}}

\begin{document}

\title[Honey, I shrunk the scientist]{Honey, I shrunk the scientist – Evaluating 2D, 3D, and VR interfaces for navigating samples under the microscope}

\author{Jan Tiemann}
\orcid{0009-0006-4588-1405}
\email{j.tiemann@hzdr.de}
\affiliation{%
  \institution{Helmholtz-Zentrum Dresden - Rossendorf}
  \institution{Technische Universit\"at Dresden}
  \institution{Center for Systems Biology Dresden}
  \institution{MPI-CBG}
  \city{Dresden}
  \state{Saxony}
  \country{Germany}
}

\author{Matthew McGinity}
\orcid{0000-0002-1179-8228}
\affiliation{%
  \institution{Technische Universit\"at Dresden}
  \institution{IXLAB}
  \city{Dresden}
  \state{Saxony}
  \country{Germany}
}

\author{Ulrik Günther}
\orcid{0000-0002-1179-8228}
\affiliation{%
  \institution{Helmholtz-Zentrum Dresden - Rossendorf}
  \city{Dresden}
  \state{Saxony}
  \country{Germany}
}

\renewcommand{\shortauthors}{Tiemann et al.}

\begin{abstract}
In contemporary biology and medicine, 3D microscopy is one of the most widely-used techniques for imaging and manipulation of various kinds of samples. Navigating such a micrometer-sized, 3-dimensional sample under the microscope---e.g. to find relevant imaging regions -- can pose a tedious challenge for the experimenter. In this paper, we examine whether 2D desktop, 3D desktop, or Virtual Reality (VR) interfaces provide the best user experience and performance for the exploration of 3D samples. We invited 12 skilled microscope operators to perform two different exploration tasks in 2D, 3D and VR and compared all conditions in terms speed, usability, and completion. Our results show a clear benefit when using VR -- in terms of task efficiency, usability, and user acceptance. Intriguingly, while VR outperformed desktop 2D and 3D in all scenarios, 3D desktop did not outperform 2D desktop.

\end{abstract}

\ccsdesc[500]{Human-centered computing~Empirical studies in HCI}
\ccsdesc[500]{Human-centered computing~Virtual reality}
\ccsdesc[300]{Human-centered computing~Graphical user interfaces}
\ccsdesc[100]{Applied computing~Systems biology}

\keywords{human-computer interaction, microscope control, virtual reality, visualization, simulation, study, desktop interface}

\begin{teaserfigure}
  \centering
  \includegraphics[width=0.96\textwidth]{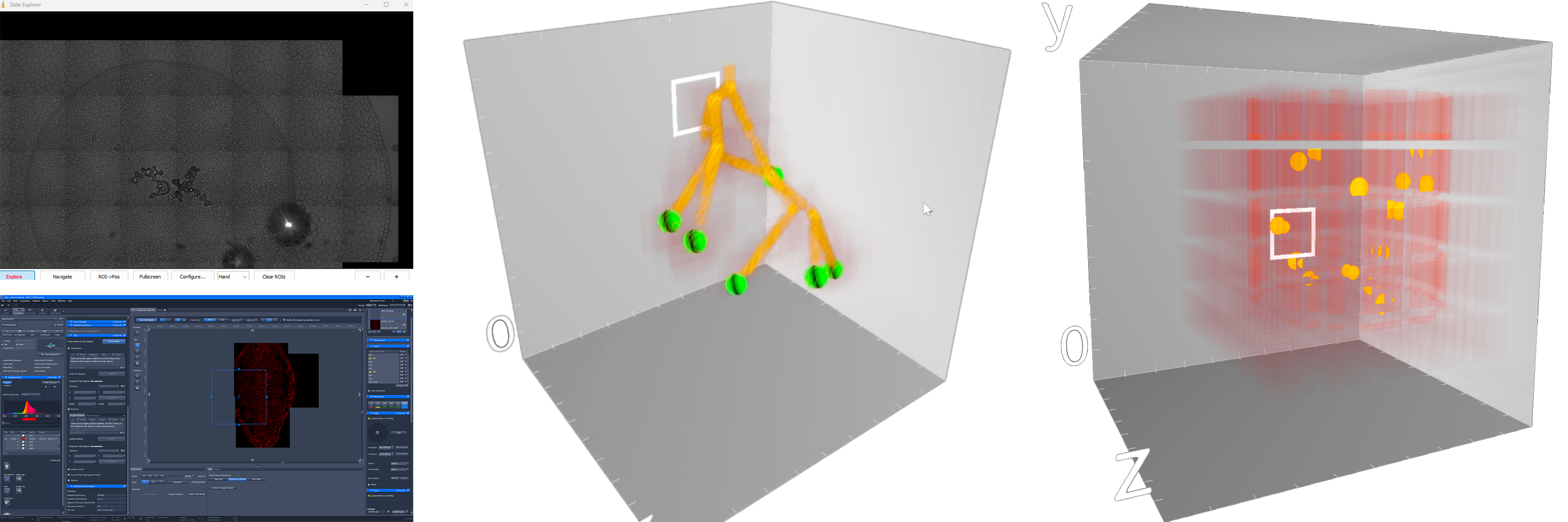}
  \caption{\textit{Left:} Two state-of-the-art microscope sample explorers with only 2D support (MicroManager SlideExplorer and Zeiss ZenBlue). \textit{Center \& Right:} Our 3D sample explorer showing both scenarios used in the study, for Axon and Tube scenarios.)}
  \Description{Two state of the art microscope sample explorer with only 2D support. Our 3D sample explorer showing both scenarios used in the study, one representing connecting and one multiple round samples in a tube.)}
  \label{fig:teaser}
\end{teaserfigure}

\received{20 February 2007}
\received[revised]{12 March 2009}
\received[accepted]{5 June 2009}

\maketitle

\section{Introduction \& Background}\label{sec:problem-description-background}

In state-of-the-art microscopy, such as fluorescence microscopy used in medical and biological research, three-dimensional volumes of image data are created by successively illuminating thin, two-dimensional slices of the sample under the microscope \cite{Reynaud:2015dx}. This process is similar to how volumes of image data are acquired in CAT scans or MRI.

\begin{figure}
    \centering
    \includegraphics[width=0.43\linewidth]{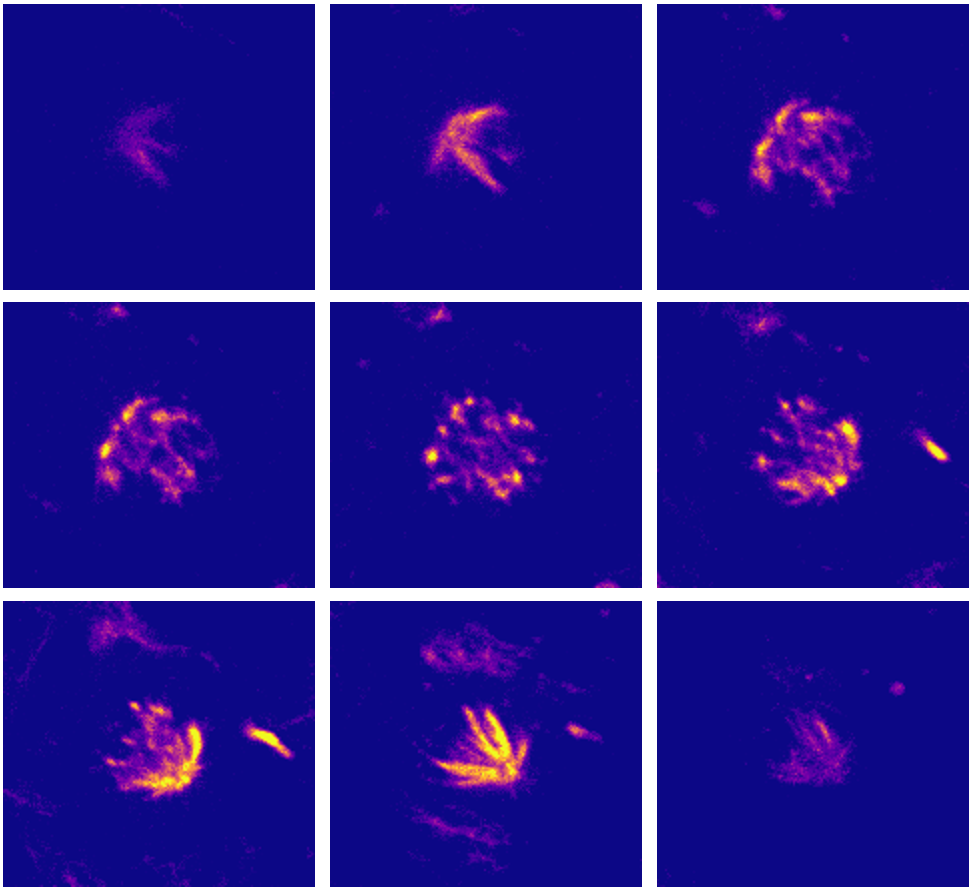}
    \includegraphics[width=0.455\linewidth]{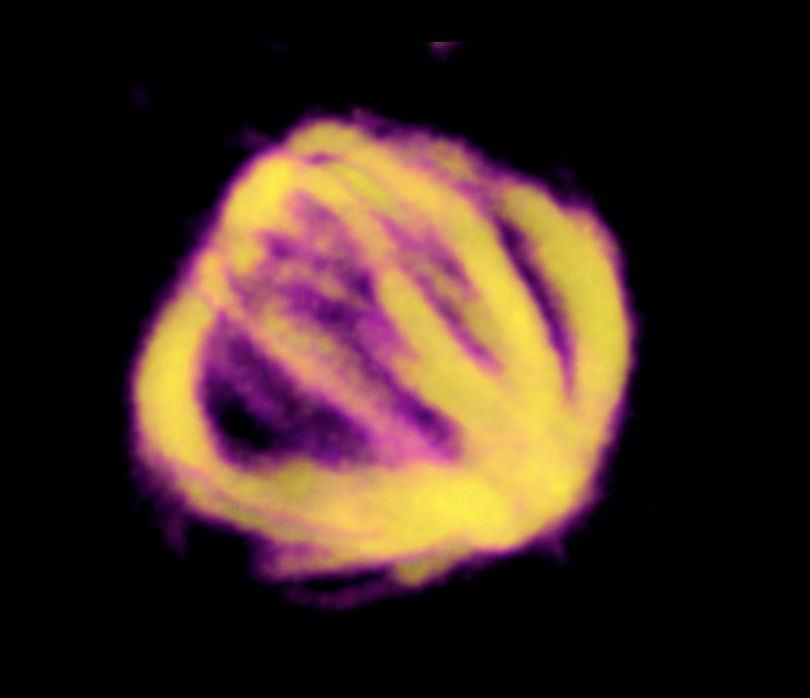}
    \caption{ Spatial understanding between 2D and 3D at the example of an image stack of a mitotic spindle which is a subcellular component orchestrating cell division. Microtubules extent from pole to pole, forming a spherical to oblong structure that facilitates the distribution of dividing chromosomes onto the daughter cells. \textit{left:} Selection of images from the image stack. \textit{right:} Volume rendering of image stack, which makes the 3-dimensional structure much clearer.}
    \Description{There are two images. The left image is a grid of nine separate, small pictures. Each picture shows a glowing, abstract shape against a very dark blue background. The first and last picture are less bright than the others. The right image shows a volume rendering of a mitotic spindle on a black background.}
    \label{fig:MitoticSpindle}
    \label{fig:MitoticSpindle3D}
\end{figure}

Despite being able to obtain volumetric 3D images, the user interface for controlling such microscopes still relies on a 2D live camera feed, displaying only one layer of the sample at a time. As the micrometer-scale samples become larger and more spatially complex, a 2D camera feed becomes insufficient for both navigation and exploration of the---potentially large---space under the microscope's objective where the sample resides, called the \emph{stage space} (compare \Cref{fig:MitoticSpindle}). Navigating this stage space is particularly challenging at the beginning of the microscopy workflow, when the sample is freshly mounted under the microscope and the regions of interest (RoI) for further imaging need to be defined. These RoIs might be the area of the sample itself, parts of the sample, or even several samples mounted for the same experiment. Conventional approaches to stage exploration are:\\

\textbf{Magnification change} -- This option requires a microscope with multiple objectives and an objective revolver. The revolver allows for minimally disruptive magnification changes by switching objectives on-the-fly. However, the objective revolver itself plus objectives with different magnifications incur a significant cost. Further, particular experiment configurations can prohibit the use of such hardware.

\textbf{Scan and stitch} -- Stitching works by obtaining images at different positions and rendering them relative to their actual spatial coordinates on the computer (an example is shown in the top left of \Cref{fig:teaser}, with the individual image boundaries clearly visible). Each image contains only a limited view of the sample. When multiple images from different X and Y positions have been obtained, stitching and rendering them together allows the researcher to get an overview over the whole sample. For flat samples, scanning the XY plane can easily be automated and executed in reasonable time (see \Cref{fig:MicroManager-SlideExplorer}). However, for samples with significant depth or more complicated structure, this approach becomes tedious, prohibitively time-consuming, and/or even damaging to the sample due to phototoxicity. Additionally, current microscope control software is usually not equipped to render 3D stitched images.\\

In both approaches, the RoIs themselves are then typically defined by either saving the coordinates of a point in the middle of the RoI, or by drawing a rectangular bounding box around it. For 2D RoIs, this poses no difficulty using keyboard and mouse. For 3D RoIs however, the process becomes significantly more involved. The situation is further complicated if there is the necessity to define multiple RoIs at the same time, or by the need for rapid acquisition---e.g.~because the examined organisms develop rapidly past the desired developmental stage, or due to fast molecular or structural decay. Improving or accelerating navigation of the 3D stage space of a microscope is therefore highly relevant for current and future experiments in both biology and medicine.

Moving towards a solution for this problem, the contributions of this work are twofold:
\begin{enumerate}
\item First, we contribute a flexible extension of \microscenery{} that enables the reproducible evaluation of nagivation tasks in 3D volumetric data based on procedurally-generated structures that emulate natural microscopic scenes, and
\item Second, we contribute a user study that compares different interfaces for navigating samples under a microscope.
\end{enumerate}

In particular, we quantitatively compare three different interfaces for microscope stage exploration: 
\begin{itemize}
    \item 2D by-slice navigation using a standard 2D WIMP (window, icon, menu, pointer) desktop interface,
    \item 3D navigation using a volume rendering of the stage space and sample together with a 2D WIMP desktop interface, and
    \item a Virtual Reality (VR) interface using volume rendering and controller-based interaction.
\end{itemize} 

We compare these different interfaces across two different tasks, with accuracy, completion time and success rate being the main benchmark. We will now first explore related work, then detail our approach and study, and finally draw conclusions on how to efficiently and accurately navigate three-dimensional samples under a microscope, as well as discuss the limitations of our approach and study. In the following we will use the term "3D desktop" to describe an WIMP interface with 3D volume rendering displayed on a regular 2D screen. Stereoscopic desktop screen setups are not part of this work.

\section{Related Work}\label{related-work}
Since we aim to evaluate a state-of-the-art 2D/3D desktop interface versus a VR interface, this section primarily focuses on studies comparing desktop interfaces with VR interfaces.

Generally, the results of publications on advantages of VR for 3D object manipulation are not as clear-cut as a daily user of VR hardware might expect, and seems to heavily depend on task and interface implementation. The contrast is probably best shown in the difference between \cite{liebersPointingItOut2024} and \cite{linImmersiveLabelingMethod2024}. Both evaluate the use of VR interfaces vs. desktop interfaces for labeling point cloud data. The findings of the first publication show that the desktop interface outperformed VR for simple tasks while there was no difference for complex tasks. The second publication on the other hand conclude strongly in favor for their VR interface over a state-of-the-art desktop interface. Further, \cite{linImmersiveLabelingMethod2024} include a second VR interface in their study, which performs far worse than the other two interfaces.

Even for simple object manipulation, conflicting research exists. \cite{mattheissNavigatingSpaceEvaluating2011,berardDidMinorityReport2009} both show that $\geq$3 Degrees of Freedom(DoF) control devices like VR-controllers could not outperform a classic computer mouse in speed or accuracy in translation and placements tasks visualized on 3D screens. 
In contrast, \cite{bellgardtWhenSpatialDevices2020} could show a 30\% speedup with equal precision of VR controllers over mouse control for a simple 3D spatial manipulation task visualized on HMD-based VR. 

One of the differences in the previous studies is the use of display types: Following the classification of \cite{mendesSurvey3DVirtual2019}, the first two employ stereoscopic window (Fishtank) setups while the third one employs a reality replacement setup. \cite{lemasurierComparing2DKeyboard2024} also employs a reality replacement setup and compares it to a screen constrained (aka. regular 2D desktop) setup for a robot teleoperation task. They found "[...]that participants had better performance in both task types when using the KBM[desktop] interface, however they experienced fewer collisions between the robot and the world in the VR interface.[...]"

Research comparing the use of VR and desktop setups for viewing and analyzing data again offers mixed results:
While \cite{horvatComparingVirtualReality2019,millaisExploringDataVirtual2018} found VR use slightly beneficial over desktop setups for analyzing complex CAD data and 3D chart data, \cite{hattabInvestigatingUtilityVR2021} in turn challenge the advantage VR offers for analyzing biomedical data for surgery planning. Participants of their study performed equally well in desktop and VR conditions for a scene understanding task and a direction estimation task using two 3D models (a liver and a pyramid).

When it comes to the analysis of volumetric data though, previous work shows a much clearer picture: \cite{choEvaluatingDepthPerception} found an overall benefit for a 3D stereo view with head tracking, where VR was perceived to enhance depth perception in volumetric data. 
\cite{lahaEffectsImmersionVisual2012} observed the most positive effects of immersion for tasks involving visually and spatially complex search for features in volumetric datasets. They suggest to use stereoscopic rendering and head tracking for tasks that involve analyzing spatially complex structures in a 3D volume.
\cite{prabhatComparativeStudyDesktop2008} investigated understanding spatial relationships -- including characterizing the general features in a volume, identifying colocated features, and reporting geometric relationships such as whether clusters of cells were coplanar. There, subjects qualitatively preferred and quantitatively performed better in immersive VR.

\cite{bueckle3DVirtualReality2021} also found a clear speed-up for using VR over 2D desktop for a block registration task, but nevertheless concluded that the speed of the desktop setup was enough for the user base to not bother with the setup and usage overhead of a VR setup.  


There are already quite a few applications using VR for biomedical applications for education, surgery planning or offline data viewing as shown by the review of \cite{venkatesanVirtualAugmentedReality2021}. Applications leveraging the advantages of VR for microscopy applications are however relatively sparse:

Ferretti et al.\cite{ferrettiVirtualRealityInterface2021} use a holographic microscope to image and manipulate sparse, floating particles or bacteria from VR. 
Yokoe et al. \cite{yokoeImmersiveMicromanipulationSystem2022} use VR to guide microinjections. Among others, Leinen et al. \cite{leinenHandControlledManipulation2016,liSEMImageBased3D2014} use VR to control the probe of a Scanning Electron Microscope. Tiemann et al. \cite{tiemannLiveInteractive3D2024} in turn focus on the use of VR to guide manipulation lasers of a fluorescence microscope. 

The systems presented so far focus on manipulation tasks. Takashina et al. \cite{takashinaVirtualRealityEnvironment2018} in turn present a virtual laboratory where microscope images can be streamed from the microscope onto a screen in virtual reality. The images can be arranged by time or as a magnification overlay if taken at different magnification. They thereby solve an exploration task using a microscope that supports magnification change as described in \Cref{sec:problem-description-background}. 
Further, Takashina et al. \cite{takashinaEvaluationNavigationOperations2019} showed that using VR for analyzing 3D microscope data can offer a significant speed-up. The task of their study was to find a RoI by rotating, translating and scaling a volume rendering of recorded image stack data. 

Beyond controlling microscopes, Usher et al. \cite{Usher:2017bda} use VR to trace neurons on large-scale volumetric data. There, VR shows a promising benefit over manually tracing neurons using 2D image stacks.

In contrast to the aforementioned works, in this work, we present a thorough comparison of 2D, 3D and VR interfaces for two tasks based on finding and navigating a complex 3D structure based on data acquired live/simulated, where the executing scientist is essentially shrunk to the scale of the sample under investigation.

\section{Study}\label{study}
In order to determine whether a 2D desktop, 3D desktop, or VR interface is best suited for sample navigation/exploration under the microscope, we conducted a quantitative user study with 12 expert microscopy users. In addition to the traditional 2D stitching interface and the VR interface, we introduced a 3D stitching-based interface that employs the same input and output modalities as the 2D system (see \Cref{desktop-3D}). This intermediate interface allowed us to assess whether performance gains stem from the use of VR itself or simply from having access to a 3D visualization.

The main objectives of the study were to assess both task performance and user acceptance of these three interfaces when addressing complex 3D exploration and navigation challenges. We designed two representative scenarios---a high-throughput search and a complex structural navigation task — to capture typical real-world use cases and to explore whether certain interfaces are better suited to specific task types.

For controlled comparison, we developed a microscope simulator as the backend for the VR microscopy system \microscenery{} (for code availability, see \Cref{code-availability}). This simulator enabled standardized scenario generation based on procedural data for all participants and ensured repeatability.

To assess user acceptance, participants were asked to complete a questionnaire in addition to performing the practical tasks. The questionnaire also helped validate the experimental tasks and interfaces by comparing them to the participants' daily laboratory experiences.

\subsection{Tasks}\label{tasks}

\begin{figure}
    \centering
    \includegraphics[width=\linewidth]{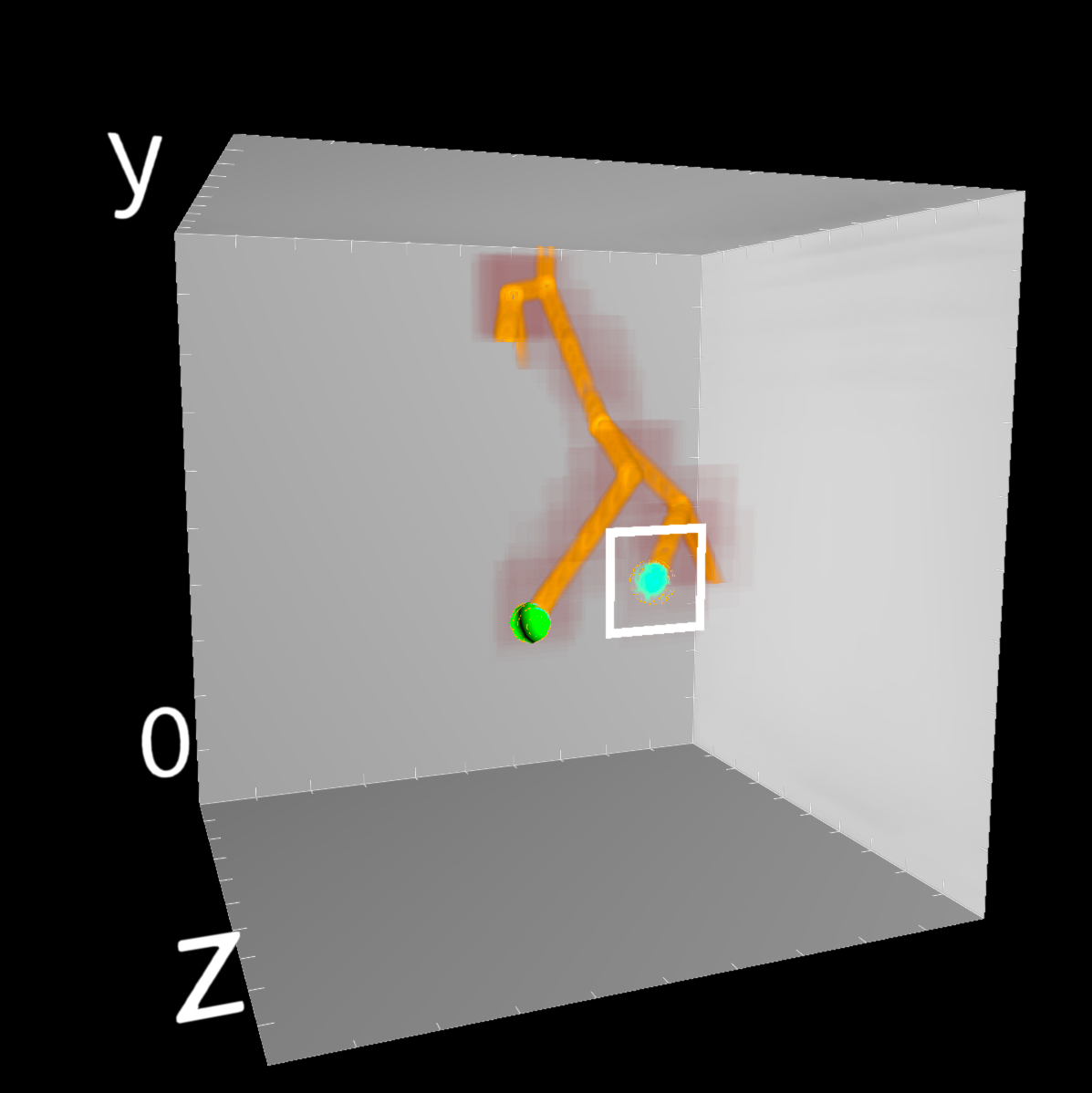}
    \caption{In-process screenshot of Axon task in 3D interface. \textit{orange:} Part of the sample structure that is already uncovered. \textit{green:} A successfully marked target. \textit{blue:} An unmarked, uncovered target. \textit{white:} The focus rectangle that is controlled by the user. Positioned to mark the next target. New microscope images are taken at that position. \textit{gray:} The box encompasses all valid positions. The sample is always fully positioned inside the box.}
    \Description{The image depicts a 3D scene, like looking into a gray box. It represents a virtual workspace used for analyzing a sample under a microscope. The scene is dominated by a structure that looks like a branching, somewhat irregular pathway, and several colored markers indicating progress in a task. The background is black, and the scene is oriented with labeled axes – ‘y’ pointing upwards, ‘o’ denoting the origin, and ‘z’ pointing forward.}
    \label{fig:3DAxon-inprocess}
\end{figure}

Both study tasks shared the same objective: participants were required to mark all regions of interest (RoIs), represented as target spheres, in each trial. To do so, they had to spatially align the center of the microscope focus rectangle with the center of a target sphere (see Figure \ref{fig:3DAxon-inprocess}). Depending on the task, targets were initially either completely or partially unscanned. To discover them, the microscope simulator would generate an image for the current focus position every 300 milliseconds, replicating the behaviour of a spinning disc confocal microscope running with a live camera. This mode of operation is usually used for navigation and exploration tasks during microscope work. The generated/acquired image would then be placed in the scene at the respective position and rendered according to each interface (described in \Cref{interfaces}). The tasks the participants were asked to perform were the following:

\textbf{High Throughput/Tube Task}\label{high-troughputtube-task} --- This task models a workflow in which researchers search for multiple, sparsely-distributed small RoIs, such as a group of fruit fly \emph{Drosophila melanogaster} embryos, inside a tubular sample holder (as described in \cite{schmiedSamplePreparationMounting2016}) or cell clusters in beaker chambers (as described in \cite{pampaloniLightSheetbasedFluorescence2015}). After mounting the tube on the microscope, the researcher must first locate the samples before the experiment can proceed. If an overview method (such as changeable magnification) is not available, the user usually starts by moving the microscope in broad strokes to find parts of the mounting hardware or sample container. Upon locating an anchor point, the user continues by navigating to areas where they suspect the targets. Then, one way to proceed is to start an automatic overview, which is still coarse, but still finer than the overview scan before. The step size of this scan should be chosen to be smaller than the size of the target object to guarantee each target is at least partially covered.

\begin{figure}
    \centering
    \includegraphics[width=0.49\linewidth]{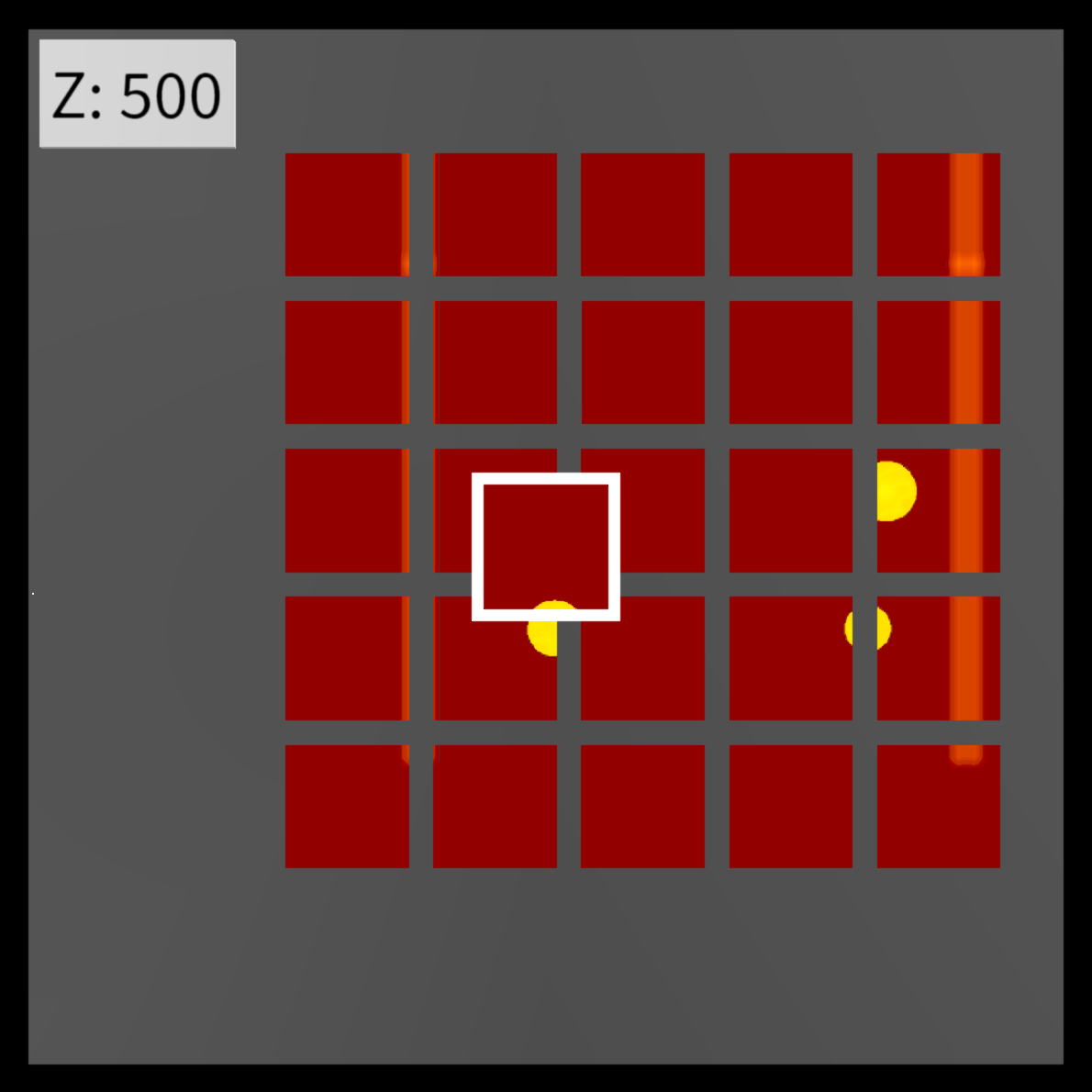}
    \includegraphics[width=0.49\linewidth]{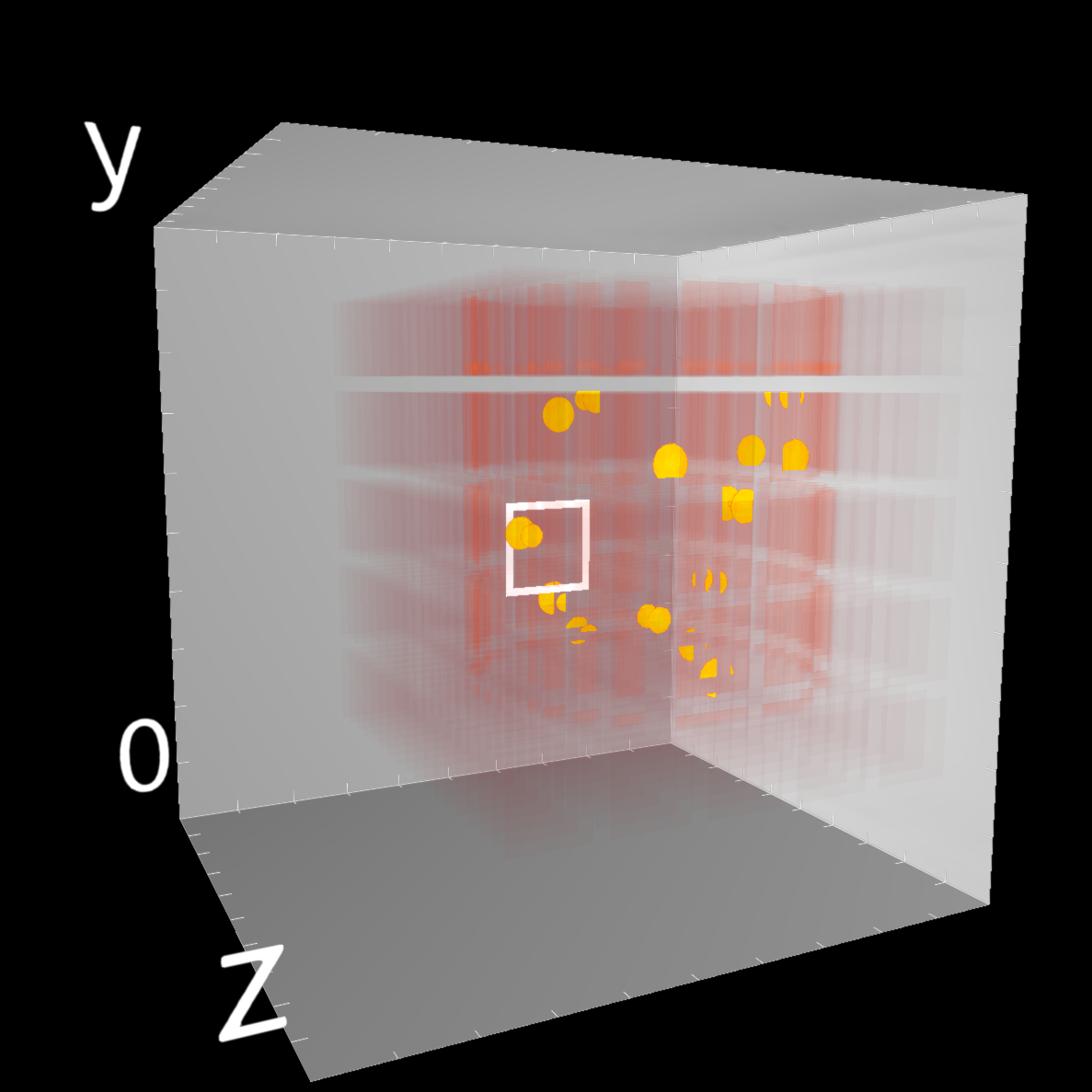}    
    \caption{Initial state of tube scenario. Partly uncovered targets are yellow while the surrounding tube structure is dark orange. \textit{Left:} in 2D interface \textit{right:} in 3D interface}
    \Description{Two sub images are present. Left: The image presents a grid-like arrangement of squares. The background is a dark gray. Most of the squares are a deep reddish-orange color, while a few have bright yellow spots. A small white square is positioned in the lower-center of the grid. A label at the top left reads “Z: 500”, indicating a depth. Right: A 3D view of the insides of a gray cube, filled with a grid of transparent red squares. A few of the squares feature yellow spots. }
    \label{fig:Tube-Inital}
\end{figure}

The state immediately following this overview scan defines the initial condition for our high throughput task (see Figure \ref{fig:Tube-Inital}). All targets are at least partially uncovered and the participant's goal is to navigate to each visible RoI and mark its position as quickly as possible. The exploration is limited to uncovering enough of the targets to estimate their centers. In a real world case, the experiment would continue around those marked RoIs.

In our simulation, visual distinction between targets and background was intentionally made clear, focusing the evaluation on navigation and control, rather than object identification. In practice, the identification of the RoI might be visually more challenging because of a low signal to noise ratio or because it requires specific domain knowledge. This is also the reason why this task can not easily be solved programmatically.

\textbf{Follow Structure/Axon Task}\label{axonfollow-structure-task} --- This task emphasizes exploration through a more complex spatial arrangement. Here, participants followed a branching, axon-like structure from a single root node to its terminal (``leaf'') nodes (comparable to \cite{Usher:2017bda,choEvaluatingDepthPerception,lahaEffectsImmersionVisual2012} but with hidden neurons instead; see Figure \ref{fig:teaser} \textit{center}). We deliberately chose this structure instead of tasks requiring exploration of a single large, contiguous sphere-like structure. Even though more real world examples for the latter exist, we anticipated it could introduce greater variability due to the influence of exploration path choices. By contrast, a branching structure guides users more deterministically along prescribed paths, making results more comparable and less dependent on random search decisions. Furthermore, we expect results from this setup to generalize to more  exploration tasks in 3D volumetric data if the user has domain knowledge.

\begin{figure}
    \centering
    \includegraphics[width=\linewidth]{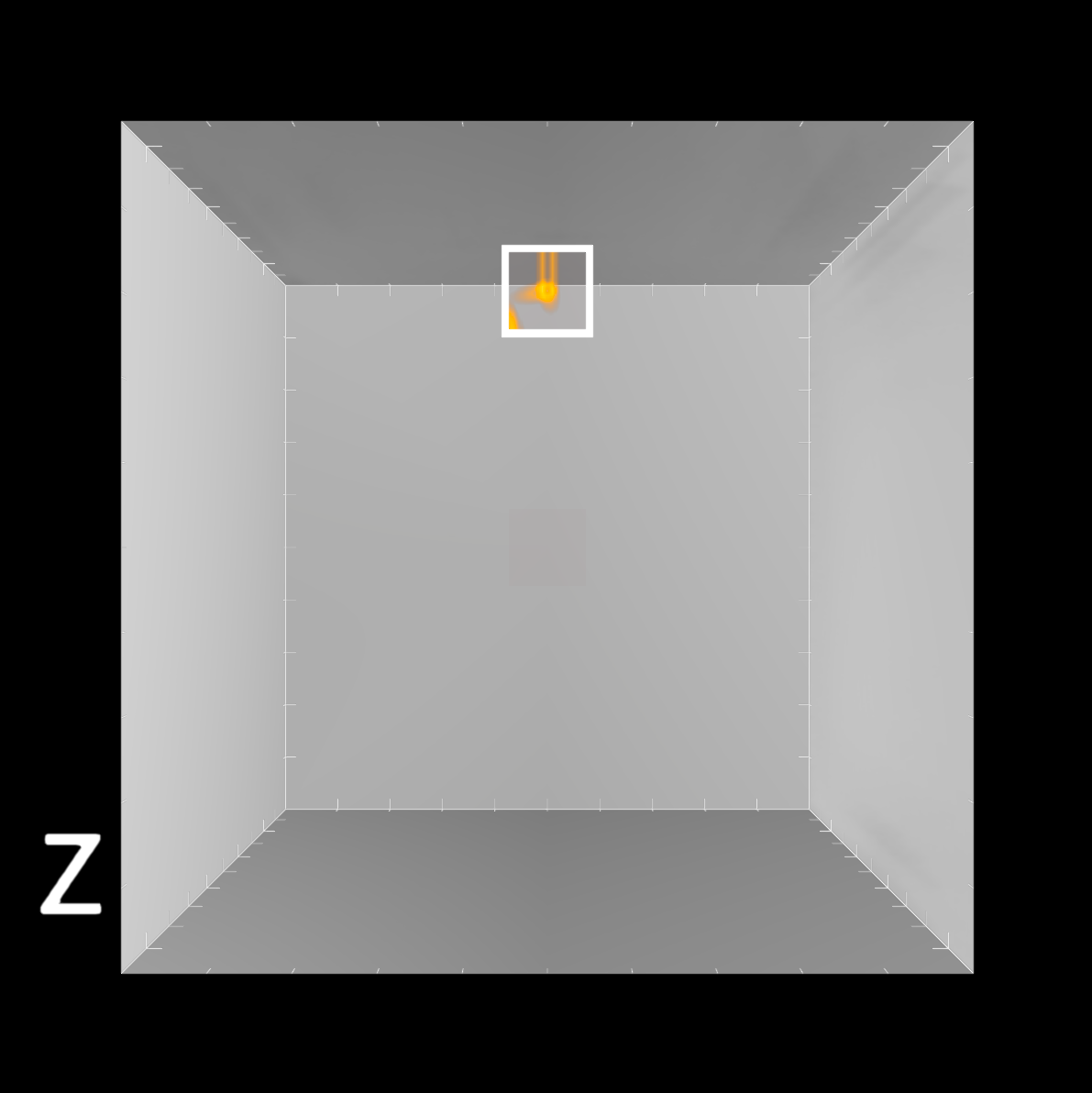}
    \caption{Initial condition for the Axon task in 3D interface. The microscope focus is placed at the root of the axon tree and thereby makes it visible. Everything else of the structure is hidden and needs to be explored by the user.}
    \Description{The image depicts a simple, empty 3D space, resembling a cube. The background is black, and the cube’s walls are a gray. The scene is oriented with labeled axes – ‘y’ pointing upwards and ‘z’ pointing forward, suggesting depth. The most striking feature is a small, glowing structure in the top center of the cube, along with a white rectangular box surrounding it.}
    \label{fig:3DAxon-Initial}
\end{figure}

Each trial began with the microscope focus at the root of the axon tree, unveiling only the root (see Figure \ref{fig:3DAxon-Initial}). The remainder of the structure was hidden, requiring participants to visually track along branches to reveal and mark all targets.

The tree structures were created using a randomized algorithm, after which a curated set of seeds was fixed for all study participants. The curation ensured non-overlapping branches and comparable path lengths, supporting consistent task difficulty and minimizing variability between trials. Each participant was exposed to the same structures generated by the same set of seeds.

\subsection{Variables and Design}\label{variables-and-design}

To systematically evaluate interface effects and task difficulty, we employed a within-subjects design with the used interface (three levels: 2D desktop, 3D desktop, and VR) as the independent variable. Additionally, we let the participants perform two different tasks (high-throughput/tube, follow-structure/axon) to cover all introduced real world scenarios. We are not analysing the effect of the tasks on the performance, instead we want to ensure that the interfaces perform well in both scenarios. This resulted in six overall experimental conditions, and each participant completed every condition twice to account for learning effects.

During all trials, we collected detailed interaction data, including the timing and outcome of each marking attempt. From these logs, we pre-defined the following dependent variables:

\begin{itemize}
\item
  \emph{mark attempt success rate}: Ratio of successful to total mark attempts within a trial.
\item
  \emph{trial completion times}: The elapsed time until all targets were marked or the three-minute time limit was reached.
\item
  \emph{targets marked per trial}: How many of the available targets in a trial were successfully marked within the allotted time.
\end{itemize}

Each trial was limited to three minutes to keep session lengths manageable. The remaining time was not shown to participants to not mount unnecessary pressure. Nevertheless, imposing a time limit also reflects realistic conditions---when exploring a sample, an experimenter must not waste time or else the examined organisms develop rapidly past the desired developmental stage or throughput demands are not met.
The order of conditions was semi-randomized for each subject, with constraints to balance interface exposure and minimize VR headset adjustments (for more detail see \Cref{process}).
 
Just like \cite{bueckle3DVirtualReality2021} had to design their experiment around the stark completion time differences between VR and desktop interfaces, we observed during analysis that a large proportion of desktop trials were ended by the time limit, while all VR trials completed successfully, compromising the comparability of the predefined metrics between VR and desktop conditions. To address this, we introduced an additional metric, the time to reach half the total targets (the \emph{50\% mark duration}), which allowed for more robust statistical comparison across all interface types (see \Cref{sec:analysis} for a detailed discussion).

\subsection{Interfaces}\label{interfaces}

To complete the tasks, participants used three different user interfaces, each based on a different metaphor for image stitching. In all cases, images acquired from the virtual microscope were placed at their spatially correct position relative to previously captured images, gradually revealing the sample structure during navigation and exploration.

\begin{figure}
    \centering
    \includegraphics[width=\linewidth]{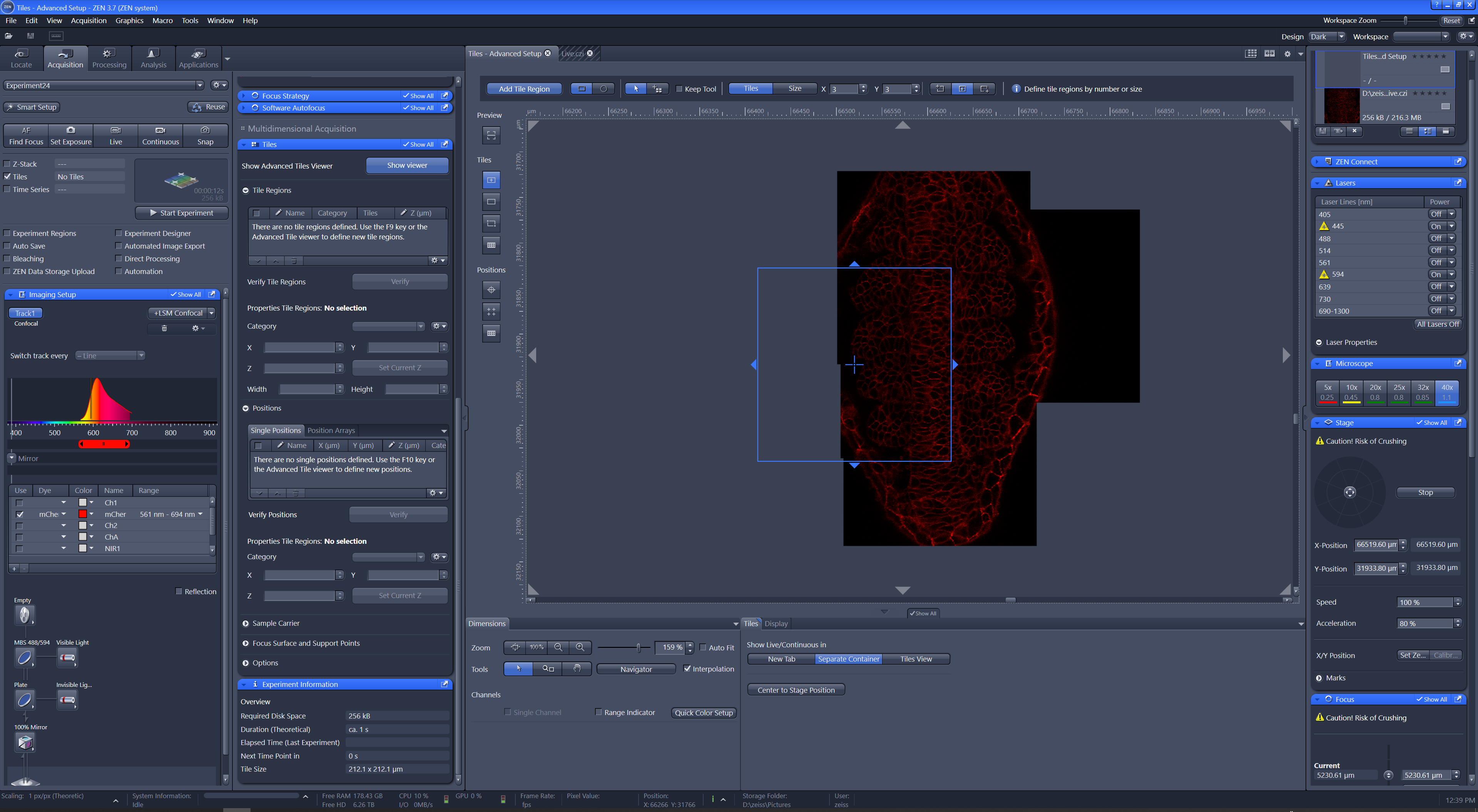}
    \caption{A screenshot of the microscope software Zeiss ZenBlue where the user is exploring a \emph{D. rerio} (zebrafish) embryo under the microscope, based on 2D slices.}
    \Description{The image is a screenshot of a complex software interface, specifically the Zeiss ZenBlue microscope software. The interface is predominantly dark blue-gray, filled with numerous buttons, sliders, and text-based panels. The central and most visually prominent element is a square image displaying a reddish, oval-shaped object. A blue rectangular box is superimposed over this image}
    \label{fig:ZenBlueTiling}
\end{figure}

\begin{figure}
    \centering
    \includegraphics[width=\linewidth]{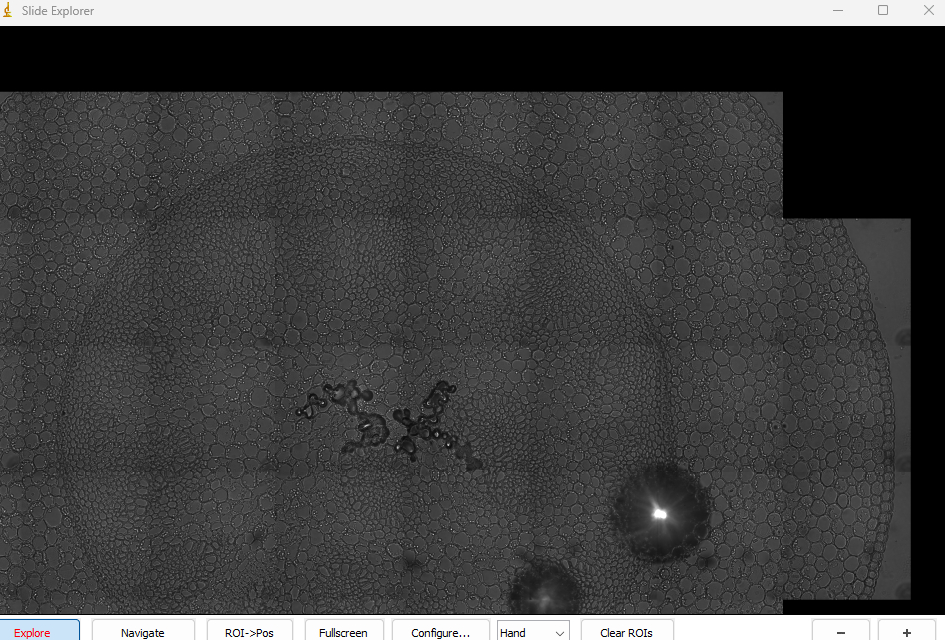}
    \caption{A screenshot of the MicroManager SlideExplorer showing \emph{Convallaria} cells.}
    \Description{The image portrays a microscopic view of plant cells, densely packed together like tiles in a mosaic. The central structure represents a different cellular arrangement, and the software interface provides tools for exploring and analyzing the image.}
    \label{fig:MicroManager-SlideExplorer}
\end{figure}

\subsubsection{2D Desktop}\label{d-desktop}

The first interface emulates state-of-the-art microscope control software, such as Zeiss ZenBlue\footnote{\href{https://www.zeiss.com/microscopy/en/products/software/zeiss-zen.html}{www.zeiss.com/microscopy/en/products/software/zeiss-zen.html}} or the open-source Micro-Manager\,\cite{edelsteinAdvancedMethodsMicroscope2014}. These platforms employ 2D image stitching, as shown in \Cref{fig:ZenBlueTiling} and \Cref{fig:MicroManager-SlideExplorer}. Newly acquired images are displayed in the correct XY plane positions to create a cohesive overview map of the sample. However, standard microscope software typically does not support direct exploration along the Z axis; for every new Z layer, a separate window must be opened, and the user receives little assistance in spatial reasoning between layers.

Our 2D desktop implementation extends this by allowing users to scroll between Z-layers using the mouse wheel, with all images oriented in a global coordinate system. This is comparable to scrolling through the images in \Cref{fig:MitoticSpindle} \textit{left}, but applied to the whole space under the microscope. Images which are taken at the same XY coordinates, but differ in the Z coordinate, are displayed at the same location when scrolling through the layers. Dragging with the left mouse button down controls the microscope focus in X and Y and pressing the middle mouse button triggers a mark attempt.

\subsubsection{3D Desktop}\label{desktop-3D}

To bridge the gap between 2D desktop control and immersive VR, we introduced a 3D desktop interface---a volumetric analog to the 2D version. Here, all images are rendered simultaneously at their corresponding XYZ positions, with transparency applied to enable volumetric perception. As a result, a basic volume rendering similar to view-aligned texture mapping emerges as more images are acquired.

\begin{figure}
    \centering
    \includegraphics[width=0.49\linewidth]{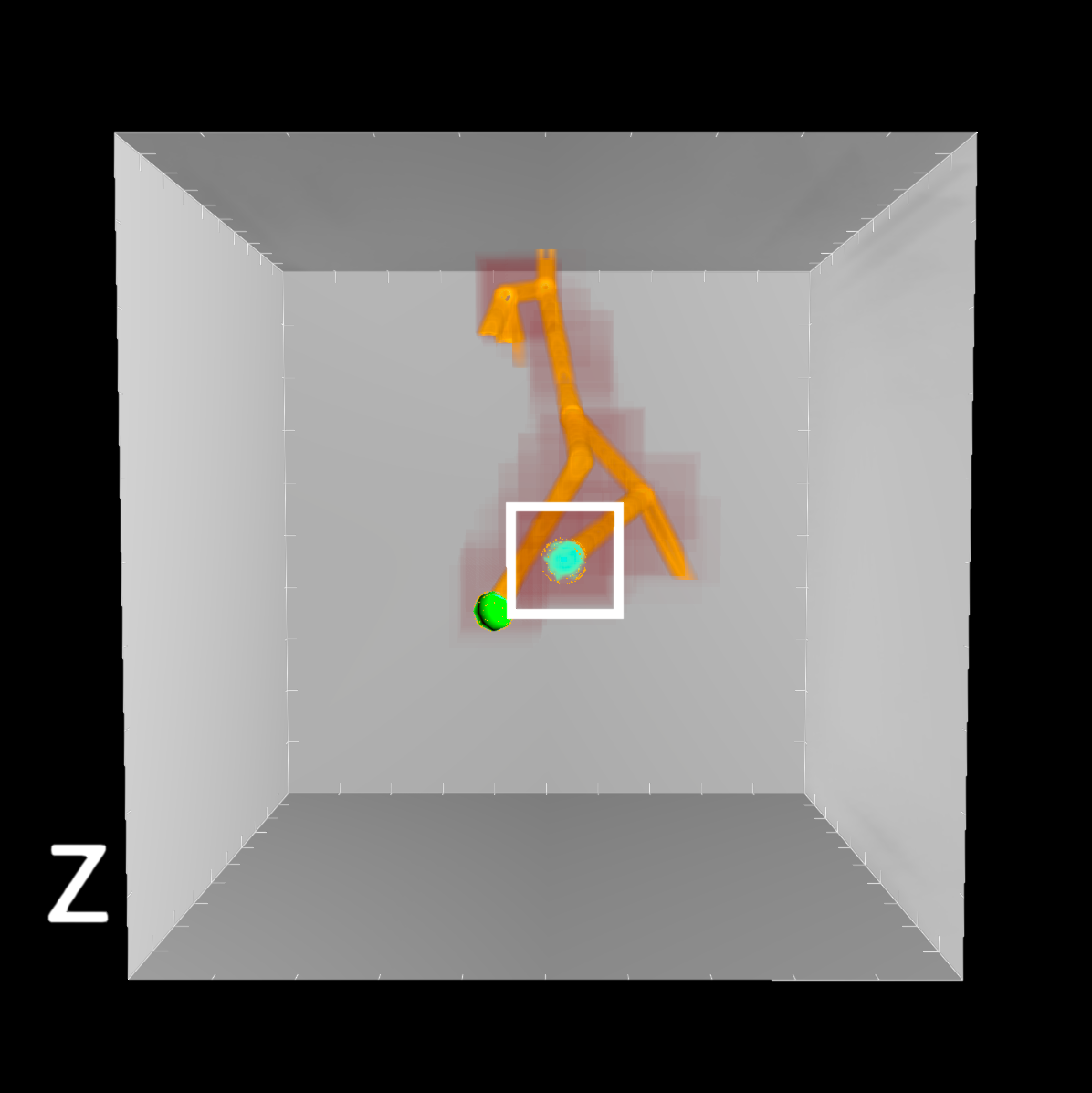}
    \includegraphics[width=0.49\linewidth]{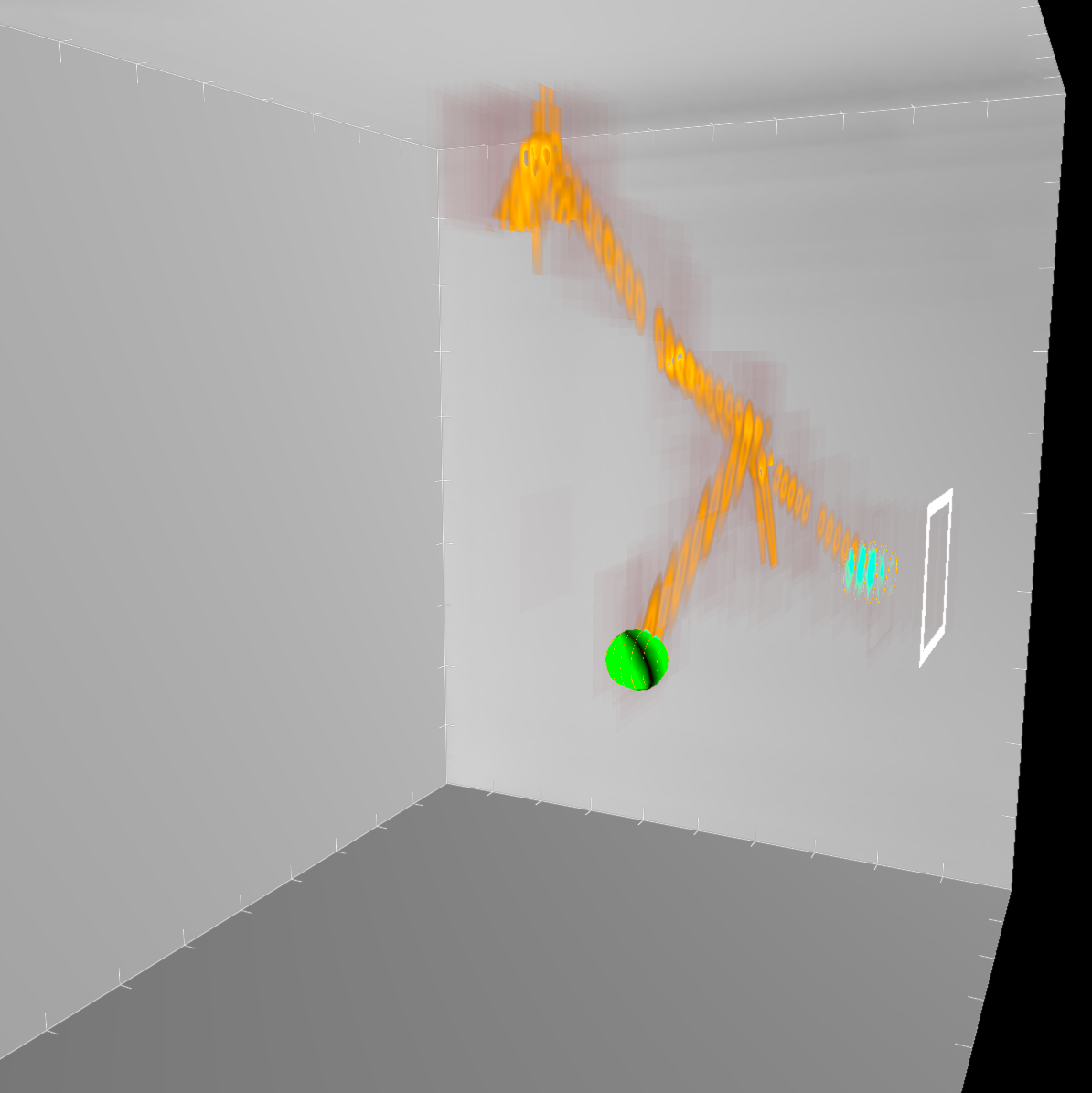}
    \caption{\textit{Left:} Visually, it looks like the focus rectangle is aligned with the target. \textit{Right:} Changing the perspective reveals that the focus is not aligned with the target along the former view direction.}
    \Description{The image is split into two side-by-side scenes, both depicting a 3D space within a gray cube. The scenes illustrate a visual illusion related to alignment and perspective. Both scenes feature a branching orange structure, green target points, and a white rectangular focus box.}
    \label{fig:3DAxon-z-perspective-optical-illusion}
\end{figure}

The interaction paradigm remains largely the same: left dragging and middle clicking controls XY focus movement and marking, respectively. In this version, the mouse wheel is used exclusively for moving the focus along the Z axis, independent of the viewing plane. Also, a dragged right click rotates the view around the current focus position. An important feature to check against are optical illusions caused by the projection to a 2D computer screen. Because of the projection from a 3D scene to a 2D view plane, it might look like the focus is aligned with a target, but in fact the alignment is only correct in the dimensions orthogonal to the view direction, and misaligned along the view direction (see \Cref{fig:3DAxon-z-perspective-optical-illusion}).

Using a three-way orthographic view, as some 3D modeling and manipulation software packages do (one view along each spatial axis is presented in three separate windows) is not a viable solution here: All obtained images are parallel to the XY plane, so all views except the one along the Z axis would just show a single pixel thick line for each obtained image. The view along Z would look like the 2D desktop interface, but with all images rendered on top each other with transparency. 

\begin{figure}
    \centering
    \includegraphics[width=\linewidth,trim={6cm 4cm 9cm 6cm},clip]{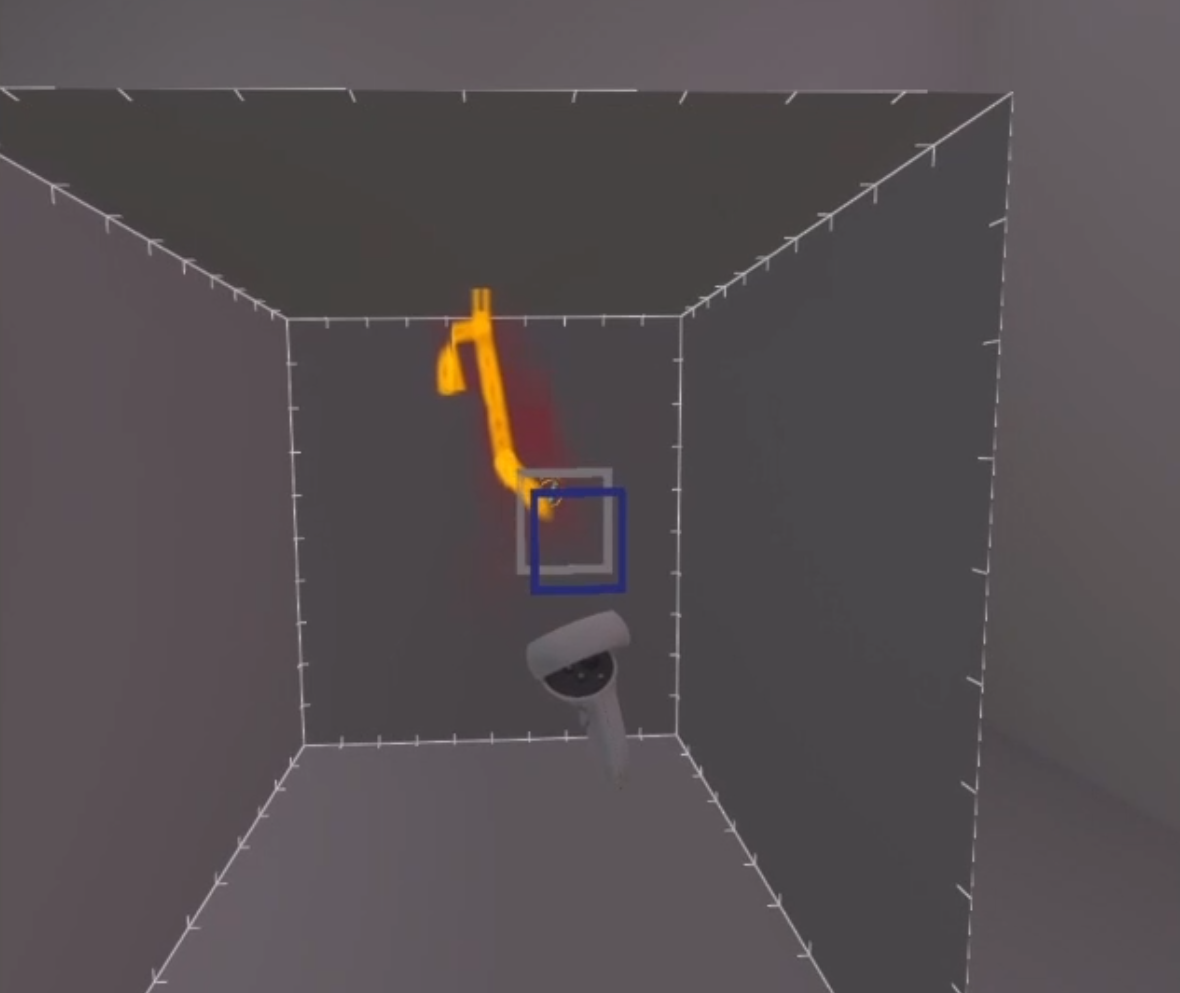}
    \caption{Screenshot of the VR interface running an Axon task. The blue rectangle is attached to the controller and marks the desired focus position. The actual focus position, marked by the white rectangle, moves towards it, adhering to a simulated hardware stage speed. }
    \Description{The image depicts a virtual reality (VR) scene inside a cube-shaped space. The scene features a bright yellow, branching structure, two rectangular boxes –--one blue and one white--– which are close to each other and a representation of a VR controller. }
    \label{fig:VRAxon}
\end{figure}

\subsubsection{VR}\label{vr}

The last interface is an immersive VR mode (see Figure \ref{fig:VRAxon}). Like the 3D desktop interface, images are rendered at their true XYZ positions; however, the scene is displayed in stereoscopic 3D on a Oculus Quest 2 HMD. The system's focus position is directly linked to the location of the user's dominant-hand VR controller, which is also used to trigger marking events. The user's viewpoint naturally follows their head position, with both hand controllers available for scene repositioning or scaling the view as needed. The interface is meant to be used in a seated position. The position of the simulation visualization can be freely moved by holding a button on the controller and moving it. The visualization then translates likewise. Also holding the grab button on both controllers and changing their distance to each other allows to scale the visualization. Rotation was not implemented. To view the visualization from another side the relative position to the user could be changed---e.g.to view the visualization from the left it can be moved to the right of the user.

\subsection{Classification}

We classified our interfaces following the taxonomy of \cite{mendesSurvey3DVirtual2019} in \Cref{tab:classification}. Since we are only translating a single always selected object, the microscope focus, some categories like scale, rotation or contact points don't apply or are set to 0 for all interfaces. The two desktop interfaces only differ in visualization, therefore they share a row in the table.

\begin{table*}[t]
    \centering
    \begin{tabular}{c|c|c|c|c|c|c|c|c}
\multicolumn{5}{c}{Environment properties}&\multicolumn{4}{|c}{Translation}\\
Technique&Display mapping&Tracking space&Separation&Hands/DOF Tracked&Mapping&CP&TD&MD \\
\hline
2D/3D desktop&screen constrained&2D Separated&\{Txy\},\{Tz\}&-&Hybrid&0&3&1 \\
VR&reality replacement&3D Co-located&\{Txyz\}&1/3&Exact&0&3&3 \\
    \end{tabular}
    \caption{Classification of the interfaces following the taxonomy of \cite{mendesSurvey3DVirtual2019}. Abbreviations: CP, number of contact points required; TD, total transformation DOFs supported; MD, minimum explicitly simultaneously controlled DOFs.}
    \label{tab:classification}
\end{table*}

\subsection{Simulating a Microscopy Environment}\label{simulator}

Our experiments were conducted using a simulation module based on our existing microscopy framework, \microscenery{}. In its standard configuration, \microscenery{} supports both direct control of physical microscope hardware and the emulation of simple microscopes using recorded image files.

For this study, we originally took three options for scenario/task generation into consideration: operating actual hardware, using previously recorded datasets, and generating simulated data. Although using real microscope hardware would maximize external validity, it was impractical due to limited access and the need for standardized, repeatable scenarios/samples. Similarly, while recorded data can enhance study realism, sourcing or generating datasets with both matched task difficulty and sufficient variability---while also not favoring participants with particular domain knowledge---proved infeasible.

As a result, we opted for simulated microscope data. Simulation enabled us to systematically create task scenarios that were both sufficiently distinct to minimize learning effects, and comparable across participants. By abstracting away domain-specific sample knowledge, we ensured that the study focused on navigation and interaction skills rather than subject-matter expertise.

Our simulator integrates seamlessly into the \microscenery{} framework via the same interface used for real hardware. The core of the simulator is a signed distance function renderer capable of generating 2D scalar images corresponding to a specified (virtual) microscope position. These simulated camera images are then processed by the framework as if they were acquired from an actual instrument. For scenario creation, experimenters can flexibly position, orient, and scale geometric primitives such as cylinders and tubes to represent sample structures, and place spheres to indicate regions of interest.

\subsection{Participants}\label{participants}

We recruited 12 participants for the study, all of whom were active in research in either cell biology or developmental biology, or working in a microscopy facility. Recruitment was carried out both via an inter-institute mailing list and through personal contacts within the local research community.

The mean participant's age was 34.4 years (median: 33, std: 8.44, min: 21, max: 55), and 6 participants identified themselves as female and 6 as male. All participants reported frequent microscope use: Seven reported at least monthly use and five daily use. Prior experience with VR was limited across the group: While all had previously used a VR headset at least once, they reported VR exposure as less frequently than monthly.

\subsection{Hardware}\label{hardware}

A Razer Blade 15 Advanced (Early 2021) notebook with a Nvidia Geforce RTX 3080 Mobile GPU (driver version 565.90) and an Intel Core i7-10875H with 2.30 GHz base clock/5.10 GHz turbo clock and 32 GiB RAM was used. The machine was running Windows 10 Pro (Version 10.0.19045). A Meta Quest 2 with a Kiwi design comfort head strap was used as head-mounted display, connected via Quest Link. Both components were connected to the network using 5\,GHz 802.11ac Wifi. A Logitech M500s mouse was used in the desktop scenarios.

\subsection{Questionnaire}\label{questionnaire}

Participants completed structured questionnaires both before and after the experimental trials (see also \Cref{adx:questionnaire}). The pre-study questionnaire collected demographic information, assessed participants' general and VR-specific experience, and evaluated their current physical and mental condition. This block assessing current condition was repeated after the study to monitor changes during participation.

The post-study questionnaire included several additional components. The validated Simulator Sickness Questionnaire (SSQ) \cite{kennedy1993} was used to evaluate any VR-induced discomfort. Participants were also asked to rate how representative they found the experimental tasks with respect to real-world microscopy work. For each interface, participants responded to repeated question blocks addressing ease of use, perceived interaction precision, overall understanding, and the likelihood of adopting the interface as a daily work tool. All quantitative items concerning interfaces and tasks were rated on a 5-point Likert scale, whereas the SSQ retained its standard 4-point scale. Each interface block concluded with an optional free-text comment field, providing space for qualitative feedback. At the end of the study, participants were additionally asked to rank the three interfaces by personal preference.

\subsection{Process}\label{process}

Prior to each session, all devices and shared equipment were disinfected. Upon arrival, participants were presented with the consent forms and completed the pre-study questionnaire. Following informed consent and the initial questionnaire, participants viewed an explanatory video outlining the study's aims and structure (see Supplementary Materials and \Cref{code-availability}).

Next, the instructor provided a hands-on introduction to the hardware and allowed participants to familiarize themselves with all interfaces in dedicated demo scenarios. Any questions were addressed at this stage to ensure that participants had a clear and adequate understanding of the tasks and interaction modes before data collection commenced.

During the study, each participant completed all six conditions twice---in a semi-randomized order designed to balance possible order effects. Two constraints governed this order: first, every participant experienced each condition once before repeating any; second, within both the first and second completion blocks, VR trials were scheduled consecutively to minimize HMD adjustments.

Upon finishing the practical tasks, participants completed the post-study questionnaire. As a token of appreciation, each participant was thanked and offered a small, previously undisclosed gift (snacks and stickers).

\subsection{Analysis}\label{sec:analysis}

\emph{Note: In all box plots, the box extends from the first quartile to the third quartile of the data. The whiskers extend from the box to the farthest data point lying within 1.5x the IQR from the box. In each individual box plot, the solid line indicates median, while the dashed line indicates the mean.}

Due to substantial performance disparities between the VR and desktop conditions, we devised an additional performance metric for statistical comparison. While all participants completed every VR trial within the allotted time, many desktop interface trials were cut short by the three-minute limit (successful trials: 3D:Axon 6/24, 3D:Tube 11/24, 2D:Axon 6/24, 2D:Tube 19/24). Therefore \emph{trial completion times} and \emph{targets marked per trial} offer no discernible performance levels for all trials. For the VR trials, \emph{trial completion times} are discernible and they all have a \emph{targets marked per trial} rate of 100\%, while for over 50\% of the desktop trials the time to finish/abort is 3:00 min, and the \emph{targets marked per trial} rate is discernible.

\begin{figure}
    \centering
    \includegraphics[width=\linewidth]{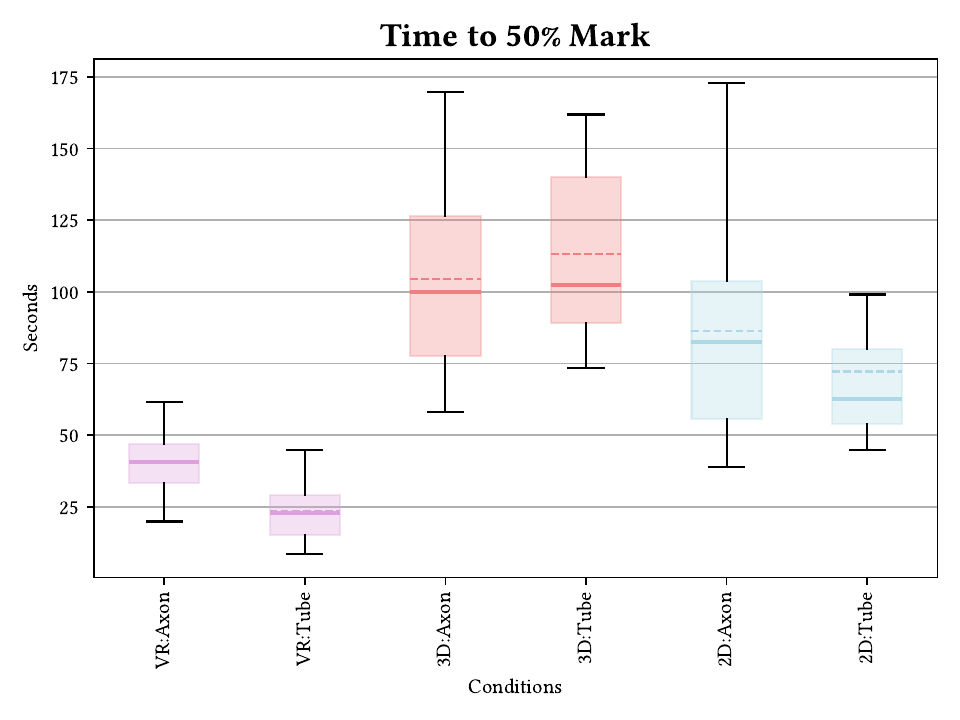}
    \caption{The time to 50\% mark plotted as a boxplot with whiskers per trial condition. The 50\% mark duration is the the time taken to successfully mark half of the predefined targets. See \Cref{sec:analysis} for a detailed explanation and \Cref{tab:50success} for numerical values.}
    \label{fig:50markchart}
\end{figure}

To address this limitation and enable meaningful statistical analysis, we introduced a new dependent variable: the \emph{50\% Mark Duration}, defined as the time taken to successfully mark half of the targets in a given trial (see \Cref{fig:50markchart}). All but four trials (three in 3D:Axon, and one in 3D:Tube) met this threshold; these outlier trials were excluded from analyses involving this metric. Notably, including these outliers would only strengthen the observed trends. Plots for the metrics \emph{completion times} and \emph{success rates} are provided in \Cref{fig:markRates} and \Cref{fig:markRates}. 

Prior to inferential analysis, we assessed normality of the data using the Shapiro-Wilk test. As most dependent variables failed to meet the criteria for normal distribution (p \textless{} 0.05), we employed non-parametric tests throughout. Specifically, Kruskal-Wallis test was used for group comparisons, including questionnaire data (as suggested by \cite{georgeschoueiryStatisticalTestsINSPECTLB2025} and \cite{nayakHowChooseRight2011}). For post-hoc analysis we used Dunns test with Bonferroni correction.

The questionnaire answers were mapped to values from 0 to 4 where a low value indicates a less favorable score depending on the question.

All analysis scripts and the full dataset are available in the supplemental material, see \Cref{code-availability}.

\section{Results}\label{results}

\subsection{Task Performance Across Interfaces}\label{task-performance-across-interfaces}

\begin{table}[t]
    \centering
    \begin{tabular}{c|c|c|c}
    Task & Interface & Median & Standard deviation \\
    \hline
    \multirow{ 3}{*}{Axon}                & VR & 41.07 & 11.46 \\
     &3D desktop & 104.37 & 31.90\\
                    &2D desktop & 86.25 & 33.81\\
    \hline
    \multirow{ 3}{*}{Tube}                & VR & 23.33 & 10.44\\
     &3D desktop & 113.12 & 28.71\\
                    &2D desktop & 72.18 & 26.16\\
    \end{tabular}
    \caption{Mean Time to 50\% Mark in seconds per task and interface.}
    \label{tab:50success}
\end{table}

The primary performance metric, the 50\% mark duration, reveals a substantial speed-up for the VR interface across both task types as seen in \Cref{tab:50success} and \Cref{fig:50markchart}. Kruskal-Wallis tests for both tasks show a clear significance level $(p < 0.001)$. The post-hoc analysis reveals that in all conditions the difference between VR and either desktop condition reached significance level $(p < 0.001)$. Strikingly, while one might anticipate improved results simply by moving from 2D to 3D desktop interaction, this was not observed: in the tube (high-throughput) task, the slight speed advantage of the 2D desktop interface versus the 3D interface even reached  significance level $(p = 0.012)$.

\begin{figure}
    \centering
    \includegraphics[width=1\linewidth]{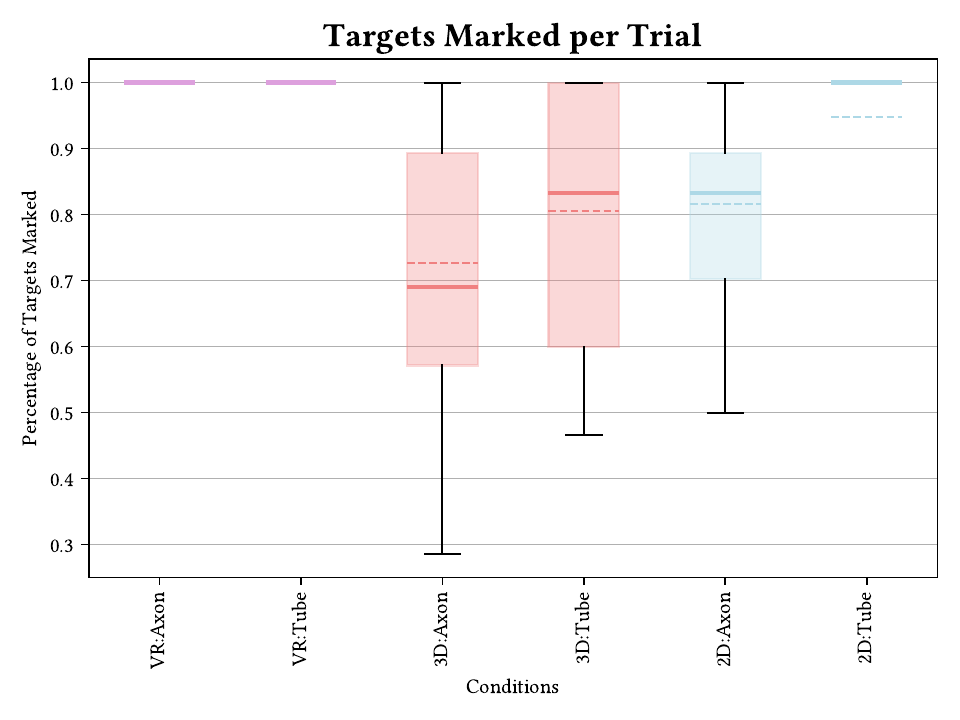}
    \caption{The percentage of marked targets per trial plotted as a boxplot with whiskers per trial condition. All VR trials, participants managed to mark all targets successfully, while the results for 2D and 3D desktop were much more varied. See \Cref{sec:analysis} for details.}
    \label{fig:markRates}
\end{figure}

\begin{figure}
    \centering
    \includegraphics[width=1\linewidth]{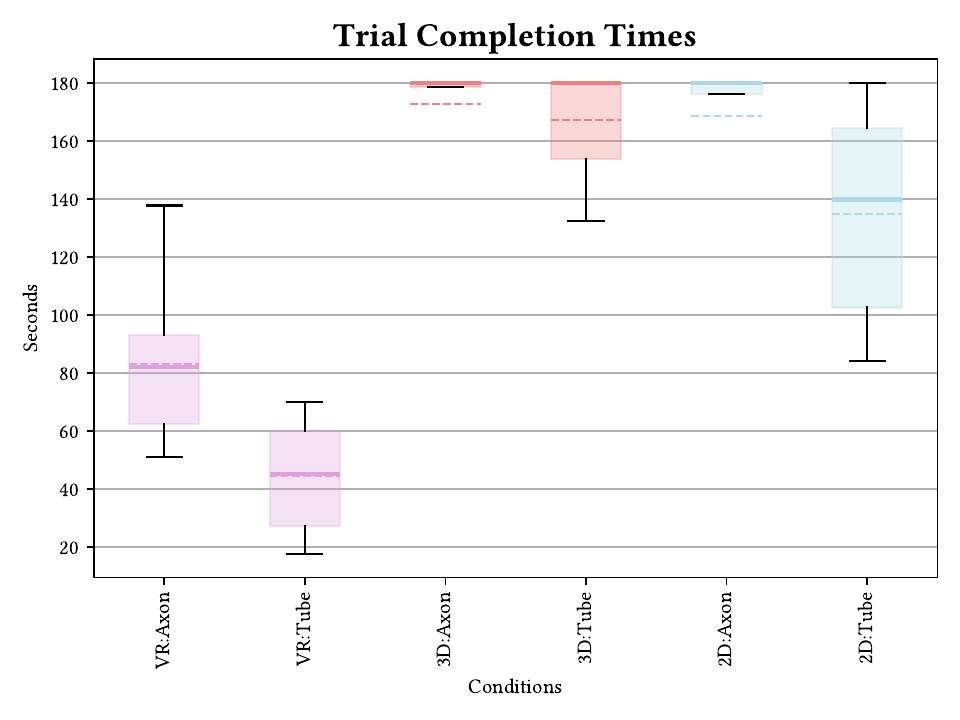}
    \caption{The time it took for participants to finish a trial or reach the 3 minute limit plotted as a boxplot with whiskers per trial condition. The most striking aspect here is that all participants managed to complete both tasks well below the time limit in the VR interface, while this was not the case for the two other interfaces, and often the time limit was hit.}
    \label{fig:trialDuration}
\end{figure}

As mentioned before, the amount of targets marked per trial and trial completion times echoed this trend: the completion rate for VR was 100\% for all conditions (\Cref{fig:markRates}), while for desktop modes the task completion time is clustered around the 3 minute time limit (\Cref{fig:trialDuration}). In other words, with respect to time limits, the tasks were perhaps too easy for the VR interface, yet too difficult for the desktop interfaces.

\begin{figure}
    \centering
    \includegraphics[width=1\linewidth]{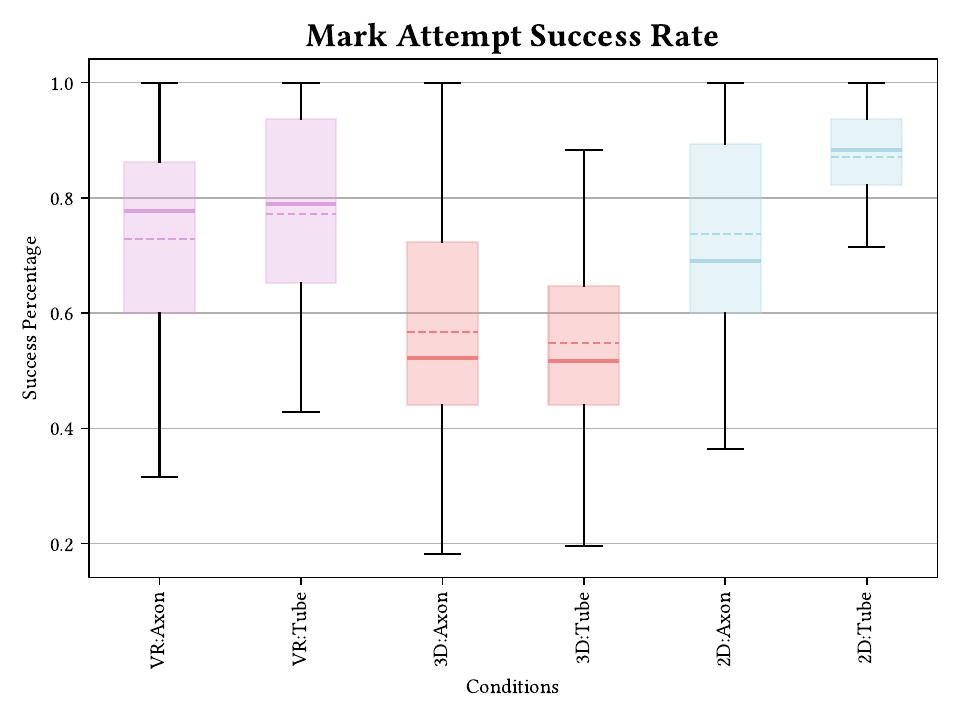}
    \caption{The ratio of successful mark attempts to the total mark attempts per trial plotted as a boxplot with whiskers per trial condition. }
    \label{fig:mark_attempt_success}
\end{figure}

\begin{table}[t]
    \centering
    \begin{tabular}{c|c|c|c}
    Task & Interface & Median & Standard deviation \\
    \hline
    \multirow{ 3}{*}{Axon}                & VR & 0.728 & 0.192 \\
     &3D desktop & 0.567 & 0.2141\\
                    &2D desktop & 0.737 & 0.198\\
    \hline
    \multirow{ 3}{*}{Tube}                & VR & 0.772 & 0.173 \\
     &3D desktop & 0.547 & 0.181\\
                    &2D desktop & 0.871 & 0.105\\
    \end{tabular}
    \caption{Ratio of successful mark attempts per task and interface.}
    \label{tab:mark_attempt_success}
\end{table}

Precision, reflected in the ratio of successful mark attempts, favor the VR and 2D desktop interfaces (\Cref{fig:mark_attempt_success} and \cref{tab:mark_attempt_success}). A mark attempt is registered when a participant presses the respective button to mark a target. It is considered successful when a new target is marked subsequently. Kruskal-Wallis tests for both tasks reach significance level (Axon:$p = 0.01$ Tube:$p < 0.001$). The following post-hoc test showed significance for all comparisons with the 3D desktop interface (Axon:$p < 0.05$ Tube:$p < 0.001$)

\subsection{Subjective Ratings and User Experience}\label{subjective-ratings-and-user-experience}

Subjective evaluations from the post-study questionnaire closely mirrored objective task results. All participants rated the VR interface as ``Enjoyable'' or higher, whereas desktop interfaces rarely exceeded a ``Neutral'' enjoyment score, with only one participant describing the 3D interface as ``Enjoyable'' (VR mean:3.750 std:0.433, 3D mean:1.500 std:0.866, 2D mean:1.250 std:0.829 Kruskal-Wallis: $p<0.001$ pos-hoc: VRvs.3D $p<0.001$, VRvs.2D $p<0.001$).

Perceived precision also favored VR, participants primarily rated VR as ``Precise'' or ``Very Precise,'' while desktop interfaces were most often described as ``Imprecise'' or ``Neutral'' (VR mean:3.333 std:0.624, 3D mean:1.250 std:0.829, 2D mean:1.833 std:0.687 Kruskal-Wallis: $p<0.001$ pos-hoc: VRvs.3D $p<0.001$, VRvs.2D $p=0.004$). 

Feedback from several participants highlighted the prevalence of depth-related perceptual errors when using desktop 3D, likely caused by optical illusions stemming from projecting a 3D scene onto a 2D screen as discussed in \Cref{desktop-3D} and shown in \Cref{fig:3DAxon-z-perspective-optical-illusion}. 

Interestingly, the highest measured precision was obtained in the 2D desktop condition, albeit participants subjectively rated this interface as less precise. This discrepancy may be due to the interface's mechanics: in 2D, if a target appears largest on the visible slice, successful marking is almost assured, but users may still perceive limitations in their spatial control.

Feeling of intuitiveness (VR mean:3.545 std:0.498, 3D mean:2.000 std:1.000, 2D mean:2.083 std:0.862 Kruskal-Wallis: $p<0.001$ pos-hoc: VRvs.3D $p=0.001$, VRvs.2D$p=0.002$),
orientation (VR mean:3.750 std:0.433, 3D mean:2.167 std:0.986, 2D mean:1.917 std:0.759 Kruskal-Wallis: $p<0.001$ pos-hoc: VRvs.3D $p=0.001$, VRvs.2D $p<0.001$),
ease of use (VR mean:3.727 std:0.445, 3D mean:1.917 std:1.115, 2D mean:2.167 std:0.799 Kruskal-Wallis: $p<0.001$ pos-hoc: VRvs.3D $p<0.001$, VRvs.2D $p=0.002$), all show the same pattern, where VR is favored over the other interfaces.

\subsection{Well-being and Simulator Sickness}\label{well-being-and-simulator-sickness}

No significant changes were reported in self-assessed physical or mental condition following the session, and Simulator Sickness Questionnaire scores remained close to baseline for all participants, suggesting minimal negative side effects from VR use.

\subsection{Validation of Study Design and Task Realism}\label{validation-of-study-design-and-task-realism}

To assess the external validity of the study, participants rated how closely the tasks reflected their day-to-day microscopy work. Both tasks were seen as realistic and relevant, with the tube (high-throughput) task most often considered a typical at least monthly occurrence (Sometimes:2, Monthly:7, Daily:3) and the axon (exploration) task aligning with scenarios encountered less often by most participants (Sometimes: 6, Monthly:4, Daily:2). All participants found the interface training and study introduction at least ``Neutral'' or ``Helpful'', with no learning difficulties reported.

Participants answered the question "How close is the interface to the capabilities of state of the art microscope software?" for each interface as follows: VR/3D Desktop/2D Desktop Worse:0/2/2 Slightly Worse:0/2/6 Comparable:0/8/4 Improvement:7/0/0 Big improvement:4/0/0.Three participants (1 in writing, 2 verbally) commented that current microscope systems typically use 3DoF controls (like joysticks and wheels) rather than mouse and keyboard.

\begin{figure}
    \centering
    \includegraphics[width=1\linewidth]{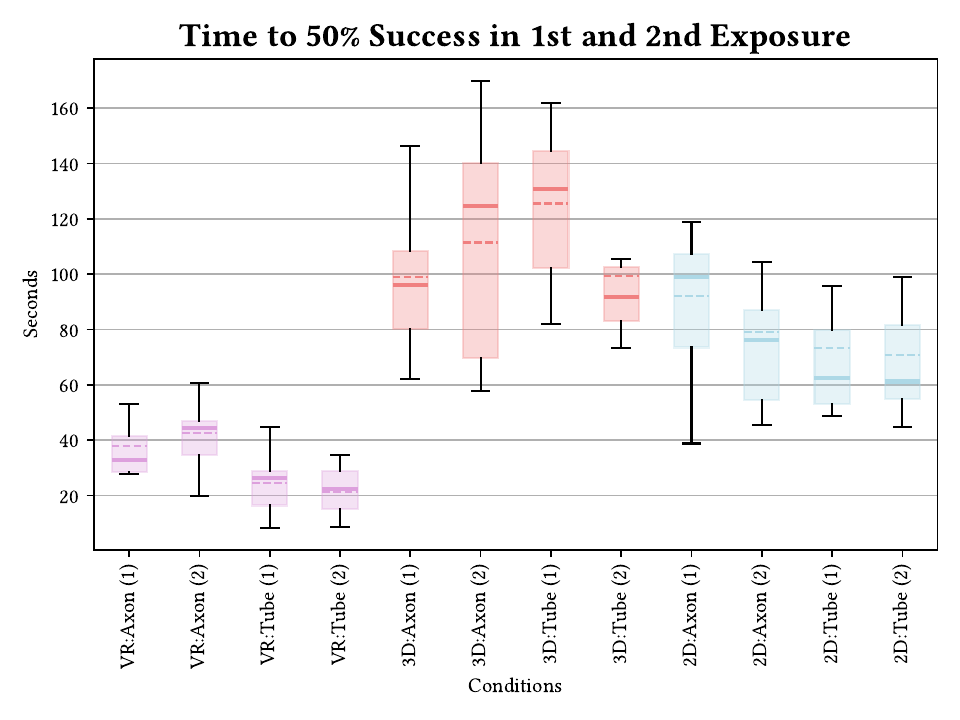}
    \caption{The 50\% mark rate for the first and second exposure of each condition plotted as a boxplot with whiskers per trial condition. No significant training effect was found in the analysis (see \Cref{sec:analysis}).}
    \label{fig:Training_Effect}
\end{figure}

\subsection{Learning Effects and Background}\label{learning-effects-and-background}

No significant relationship was observed between prior experience with 3D software and performance, nor was there evidence of systematic improvements between a participant's first and second exposure to any condition as seen in \Cref{fig:Training_Effect}.

\section{Limitations}\label{limitations}

Recruiting expert microscope users proved to be a significant challenge and left us with a sample size that endangers the external validity even for this subgroup. We still think this constraint was good decision because it allowed us to presumably observe a performance level in the 2D interface that resembles real world performance as close as possible.

Further, a participant count of this size would have benefited from rigorous counterbalancing which is lacking, as we expected a larger participant count. Only a weak form of balancing was applied (see \Cref{process}).

Participants reported that they use a joystick + Z-wheel interface in their regular microscopy work (see \Cref{validation-of-study-design-and-task-realism}). Incorporating such a device may improve the performance of the desktop settings.

Even though there are notable limitations, the recorded effect size is (analysis breaking) large. Therefore, we are still confident in the conclusions following this section.

\section{Discussion and Conclusion}\label{discussion-and-conclusion}

Our findings clearly demonstrate that VR is a viable and highly effective interaction method for 3D navigation and exploration in microscopy environments. VR consistently outperformed both state-of-the-art 2D interfaces and 3D desktop interfaces, not only in quantitative performance measures but also in user acceptance and subjective ratings. Although we could not perfectly replicate all features of current commercial systems, we believe our benchmarks are sufficiently close to support robust conclusions. Moreover, the effect sizes observed were so pronounced that traditional statistical analysis was at times limited by ceiling effects, further underscoring the superiority of VR over desktop-based approaches for these tasks.

One notable finding is that 3D desktop interfaces displayed in 2D screens did not represent a suitable substitute for immersive VR in dealing with spatially complex microscopy workflows. In fact, when VR is not a practical option, our results suggest that it may be preferable for developers of microscope control software to retain a 2D interface rather than attempt to extend existing desktop systems to three dimensions, especially given the additional risk of perceptual errors in depth judgment found in our study. One explanation for 2D desktop and VR outperforming 3D desktop could be that the participants in the study were very much used to a 2D interface from their daily work, yet were intuitively able to use the VR interface as it relies less on metaphors and more on natural movements and behaviours.

We are now collaborating with biologists to apply our interface to real-world biological photomanipulation experiments, easing experiments like \cite{boutillonDeepSpatiallyControlled2021} and providing a further opportunity to validate our approach in practice. The marking of targets for photo-manipulation is closely analogous to the RoI identification tasks explored in this study, and our previous qualitative research \anon[redacted for anonymous review]{\cite{tiemannLiveInteractive3D2024}} also supports the transferability of these findings to experimental workflows. For future research, we plan to investigate whether differences in input modalities, such as mouse and keyboard versus hardware controls like wheels and joysticks or Depth-Adaptive Cursor techniques like \cite{zhouInDepthMouseIntegrating2022}, have a measurable impact on performance and user experience. Additionally, a display condition with a fishtank 3D screen could be intriguing, especially if they can match the performance of HMD VR, as on could hypothesize from the results from studies such as \cite{mattheissNavigatingSpaceEvaluating2011,berardDidMinorityReport2009}.

We are in the process of porting our software to Augmented Reality (AR) for use in aforementioned experiments, in order to leverage the benefits of easy context switches shown by \cite{wentzelSwitchSpaceUnderstandingContextAware2024}. Using AR instead of VR will furthermore increase the viability in a lab environment by reducing the risk of accidentally destroying expensive lab equipment, due to the user being oblivious of their surroundings when fully immersed in VR. Also, manual adjustments to sample holders or laser equipment might still be performed while wearing AR hardware. Furthermore, the complexity and breadth of configuration options offered by modern microscope software (see Figure \ref{fig:ZenBlueTiling}) make a case for a hybrid approach: leveraging the established desktop interface for detailed configuration, while using spatial interaction methods, such as those presented here, specifically where they provide the greatest benefit. In this way, mixed reality environments can integrate the best features of both traditional and immersive interfaces, ultimately supporting more efficient and intuitive workflows for complex microscopy tasks.

Overall, we conclude that shrinking the scientist---enabling them to explore their samples under the microscope as a life-sized representation in VR---is a viable alternative to traditional 2D interfaces and provides tangible benefits in terms of efficiency and ergonomics, while maintaining the accuracy of a traditional interface. 

\section*{Code and Data Availability}\label{code-availability}

The software used in the study, including the 2D desktop, 3D desktop, and VR interfaces is available from \url{https://anonymous.4open.science/r/DFD7DFD7}. The data acquired in the user study as well as the analysis code is available in the supplemental material to the paper. The supplemental video to the paper includes the video that was shown to all participants prior to the beginning of the evaluation session.

\section*{Statement on the use of AI}
LLMs were used strictly only for improving formulations in the text, and for improving the visual appeal of the plots. No data analysis was performed or influenced via AI.

\begin{acks}
To Ivo F. Sbalzarini for providing resources and mailing list access.
To Samuel Pantze for thorough and fast reviews.
To GWDG High Performance Computing \cite{doosthosseiniChatAISeamless2024} for providing OpenAI’s ChatGPT-4.1 that was used for assistance in enhancing the wording and sentence structure of this publication.
\end{acks}

\bibliographystyle{ACM-Reference-Format}
\bibliography{tex/bibliography.bib}

\clearpage
\appendix
{\Huge \noindent Appendix}

\clearpage
\section{Cheat sheet} \label{adx:cheatsheet}
This cheat sheet was part of the study environment and served as a low barrier reminder of the controls for the participants.
\includepdf[pages=-,pagecommand={},width=\textwidth]{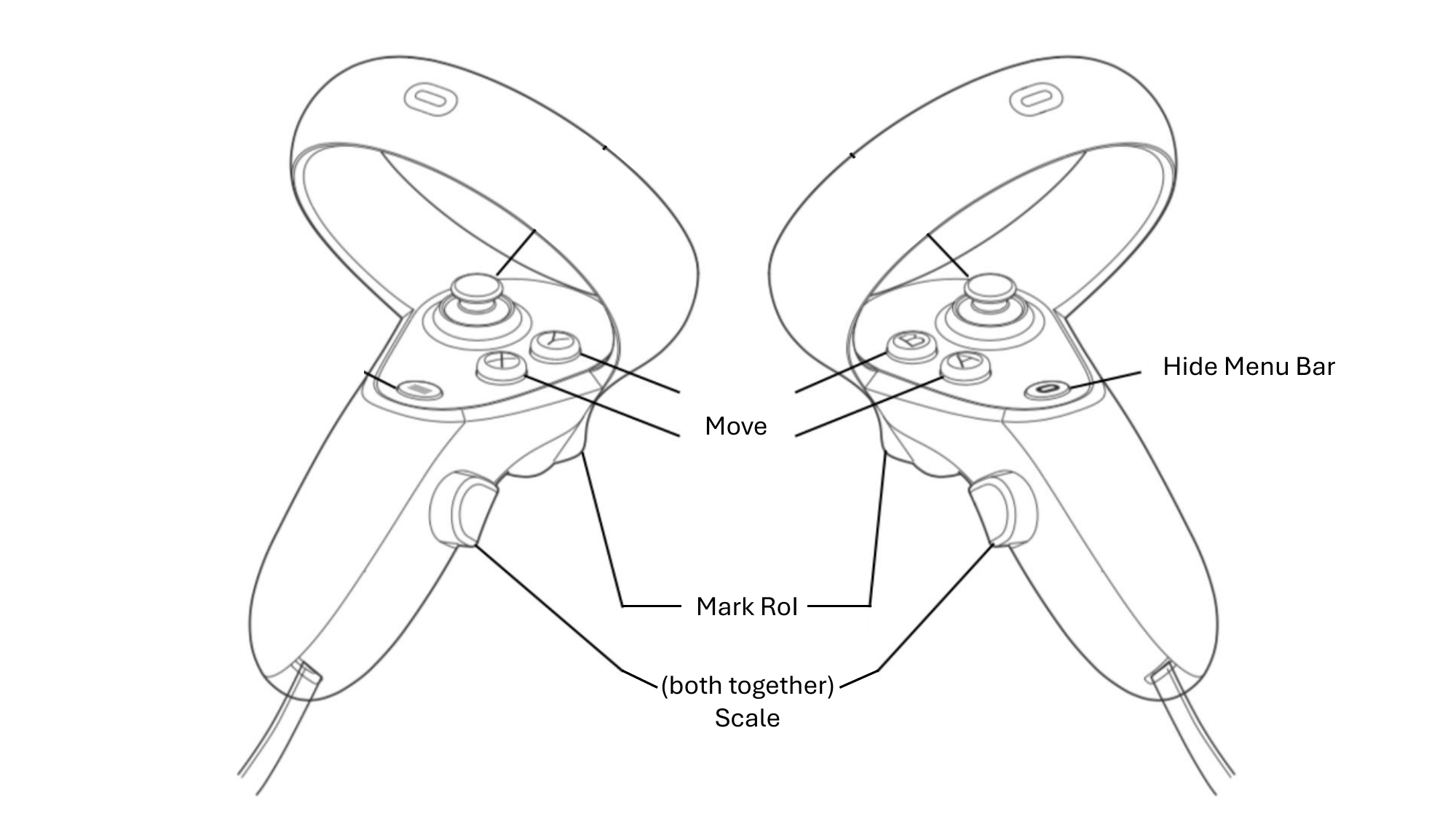}

\clearpage
\section{Questionnaire} \label{adx:questionnaire}


\section*{Supplementary material}

# B  Questionnaire

# Stage Viewer Study Questionaire

## About You

Age

Gender

| Female | Male | Other |
|--------|------|-------|
| ☐ | ☐ | ☐ |

Dominant Hand

| Left | Right | Other |
|------|-------|-------|
| ☐ | ☐ | ☐ |

Do you wear glasses or corrective lenses?

| No | Yes | If yes, please elaborate |
|----|-----|--------------------------|
| ☐ | ☐ | |

Are you wearing any right now?

| No | Yes | If yes, please specify |
|----|-----|------------------------|
| ☐ | ☐ | |

Do you have any color vision impairments, e.g., partial or full color blindness?

| No | Yes | If yes, please elaborate |
|----|-----|--------------------------|
| ☐ | ☐ | |

Do you have any impairments in spatial perception?

| No | Yes | If yes, please elaborate |
|----|-----|--------------------------|
| ☐ | ☐ | |

Do you have any motor skill impairments?

| No | Yes | If yes, please elaborate |
|----|-----|--------------------------|
| ☐ | ☐ | |

## Pre-study questions

### Prior experiences

Do you have any prior experience with VR applications or games?

| Never | Once | Sometimes | Monthly | Daily |
|-------|------|-----------|---------|-------|
| ☐ | ☐ | ☐ | ☐ | ☐ |

Have you used computer-based VR headsets before? Examples ~~for such devices~~ are Oculus Rift, HTC Vive, etc.

| Never | Once | Sometimes | Monthly | Daily |
|-------|------|-----------|---------|-------|
| ☐ | ☐ | ☐ | ☐ | ☐ |

Did you enjoy the experience?

| Not at all | It was bad | Neutral | A bit | Very much | Not applicable |
|------------|------------|---------|-------|-----------|----------------|
| ☐ | ☐ | ☐ | ☐ | ☐ | ☐ |

Have you used smartphone-based VR headsets before? Examples are google cardboard vr, Zeiss VR ONE Plus, etc.

| Never | Once | Sometimes | Monthly | Daily |
|-------|------|-----------|---------|-------|
| ☐ | ☐ | ☐ | ☐ | ☐ |

Did you enjoy the experience?

| Not at all | It was bad | Neutral | A bit | Very much | Not applicable |
|------------|-----------|---------|-------|-----------|----------------|
| ☐ | ☐ | ☐ | ☐ | ☐ | ☐ |

Have you used standalone VR headsets before? Examples are Oculus Go, or Apple Vision Pro

| Never | Once | Sometimes | Monthly | Daily |
|-------|------|-----------|---------|-------|
| ☐ | ☐ | ☐ | ☐ | ☐ |

Did you enjoy the experience?

| Not at all | It was bad | Neutral | A bit | Very much | Not applicable |
|------------|-----------|---------|-------|-----------|----------------|
| ☐ | ☐ | ☐ | ☐ | ☐ | ☐ |

How much experience do you have with 3D software? (e.g. Tracking, Games, 3D-Graphics, CAD)

| Never | Once | Sometimes | Monthly | Daily |
|-------|------|-----------|---------|-------|
| ☐ | ☐ | ☐ | ☐ | ☐ |

How often do you use light sheet or confocal microscopes?

| Never | Once | Sometimes | Monthly | Daily |
|-------|------|-----------|---------|-------|
| ☐ | ☐ | ☐ | ☐ | ☐ |

How tired are you?

| Not at all | A bit | Some | Very | Very much |
|------------|-------|------|------|-----------|
| ☐ | ☐ | ☐ | ☐ | ☐ |

How good is your ability to concentrate at the moment?

| Not at all | A bit | Some | Very | Very much |
|------------|-------|------|------|-----------|
| ☐ | ☐ | ☐ | ☐ | ☐ |

How motivated are you?

| Not at all | A bit | Some | Very | Very much |
|------------|-------|------|------|-----------|
| ☐ | ☐ | ☐ | ☐ | ☐ |

Do you have a headache?

| Not at all | A bit | Some | Very | Very much |
|------------|-------|------|------|-----------|
| ☐ | ☐ | ☐ | ☐ | ☐ |

Do you have dry or aching eyes?

| Not at all | A bit | Some | Very | Very much |
|------------|-------|------|------|-----------|
| ☐ | ☐ | ☐ | ☐ | ☐ |

Do you feel nauseous?

| Not at all | A bit | Some | Very | Very much |
|------------|-------|------|------|-----------|
| ☐ | ☐ | ☐ | ☐ | ☐ |

# Stage Viewer Study Questionaire

## Post study questions

### Current Condition

How tired are you?

| Not at all | A bit | Some | Very | Very much |
|:---:|:---:|:---:|:---:|:---:|
| ☐ | ☐ | ☐ | ☐ | ☐ |

How good is your ability to concentrate at the moment?

| Not at all | A bit | Some | Very | Very much |
|:---:|:---:|:---:|:---:|:---:|
| ☐ | ☐ | ☐ | ☐ | ☐ |

How motivated are you?

| Not at all | A bit | Some | Very | Very much |
|:---:|:---:|:---:|:---:|:---:|
| ☐ | ☐ | ☐ | ☐ | ☐ |

Do you have a headache?

| Not at all | A bit | Some | Very | Very much |
|:---:|:---:|:---:|:---:|:---:|
| ☐ | ☐ | ☐ | ☐ | ☐ |

Do you have dry or aching eyes?

| Not at all | A bit | Some | Very | Very much |
|:---:|:---:|:---:|:---:|:---:|
| ☐ | ☐ | ☐ | ☐ | ☐ |

Do you feel nauseous?

| Not at all | A bit | Some | Very | Very much |
|:---:|:---:|:---:|:---:|:---:|
| ☐ | ☐ | ☐ | ☐ | ☐ |

### Simulator Sickness

Do you experience any of the following symptoms? How badly?

| Symptom | None at all | Mild | Medium | Very |
|---|:---:|:---:|:---:|:---:|
| General discomfort | ☐ | ☐ | ☐ | ☐ |
| Fatigue | ☐ | ☐ | ☐ | ☐ |
| Headache | ☐ | ☐ | ☐ | ☐ |
| Eyestrain | ☐ | ☐ | ☐ | ☐ |
| Difficulty focussing | ☐ | ☐ | ☐ | ☐ |
| Increased salivation | ☐ | ☐ | ☐ | ☐ |
| Sweating | ☐ | ☐ | ☐ | ☐ |
| Nausea | ☐ | ☐ | ☐ | ☐ |
| Difficulty concentrating | ☐ | ☐ | ☐ | ☐ |
| Fullness of head | ☐ | ☐ | ☐ | ☐ |
| Blurred vision | ☐ | ☐ | ☐ | ☐ |
| Dizzy with open eyes | ☐ | ☐ | ☐ | ☐ |
| Dizzy with closed eyes | ☐ | ☐ | ☐ | ☐ |
| Vertigo | ☐ | ☐ | ☐ | ☐ |
| Stomach awareness | ☐ | ☐ | ☐ | ☐ |
| Burping | ☐ | ☐ | ☐ | ☐ |

## Interfaces

In the following section there will be  questions about each interface.

The following questions concern the **2D Desktop** interface.

How close is the interface to the capabilities of state of the art microscope software?

| Worse | Slightly Worse | Comparable | Improvment | Big improvment |
|:---:|:---:|:---:|:---:|:---:|

☐          ☐          ☐          ☐          ☐

How enjoyable was your experience?

Very Unenjoyable    Unenjoyable    Neutral    Enjoyable    Very Enjoyable
☐          ☐          ☐          ☐          ☐

How helpful was the interface when performing the Many-Targets/Tube task?

Very Unhelpful    Unhelpful    Neutral    Helpful    Very Helpful
☐          ☐          ☐          ☐          ☐

How helpful was the interface when performing the Follow-Structure/Tree task?

Very Unhelpful    Unhelpful    Neutral    Helpful    Very Helpful
☐          ☐          ☐          ☐          ☐

How easy was it to orient yourself in the stage space?

Very Difficult    Difficult    Neutral    Easy    Very Easy
☐          ☐          ☐          ☐          ☐

How easy was it to move the focus to mark targets?

Very Difficult    Difficult    Neutral    Easy    Very Easy
☐          ☐          ☐          ☐          ☐

How natural and intuitive did it feel to use the desktop interface?

Very unintuitive    Unintuitive    Neutral    Intuitive    Very intuitive
☐          ☐          ☐          ☐          ☐

How difficult was it to learn how to use the interface?

Very Difficult    Mildly Difficult    Neutral    Easy    Very Easy
☐          ☐          ☐          ☐          ☐

Was the introduction helpful?

Not at all    Unhelpful    Neutral    Helpful    Very helpful
☐          ☐          ☐          ☐          ☐

How precise did the interaction feel?

Very imprecise    Imprecise    Neutral    Precise    Very Precise
☐          ☐          ☐          ☐          ☐

Could you imagine a microscopist using this interface?

Never    If they have to    Neutral    Sometimes    Everytime
☐          ☐          ☐          ☐          ☐

Here you can write any comment you may have about the 2D Desktop interface.

☐

The following questions concern the **3D Desktop** interface.

How close is the interface to the capabilities of state of the art microscope software?

Worse    Slightly Worse    Comparable    Improvment    Big improvment
☐          ☐          ☐          ☐          ☐

How enjoyable was your experience?

Very Unenjoyable    Unenjoyable    Neutral    Enjoyable    Very Enjoyable
☐          ☐          ☐          ☐          ☐

How helpful was the interface when performing the Many-Targets/Tube task?

| Very Unhelpful | Unhelpful | Neutral | Helpful | Very Helpful |
| ☐ | ☐ | ☐ | ☐ | ☐ |

How helpful was the interface when performing the Follow-Structure/Tree task?

| Very Unhelpful | Unhelpful | Neutral | Helpful | Very Helpful |
| ☐ | ☐ | ☐ | ☐ | ☐ |

How easy was it to orient yourself in the stage space?

| Very Difficult | Difficult | Neutral | Easy | Very Easy |
| ☐ | ☐ | ☐ | ☐ | ☐ |

How easy was it to move the focus to mark targets?

| Very Difficult | Difficult | Neutral | Easy | Very Easy |
| ☐ | ☐ | ☐ | ☐ | ☐ |

How natural and intuitive did it feel to use the desktop interface?

| Very unintuitive | Unintuitive | Neutral | Intuitive | Very intuitive |
| ☐ | ☐ | ☐ | ☐ | ☐ |

How difficult was it to learn how to use the interface?

| Very Difficult | Mildly Difficult | Neutral | Easy | Very Easy |
| ☐ | ☐ | ☐ | ☐ | ☐ |

Was the introduction helpful?

| Not at all | Unhelpful | Neutral | Helpful | Very helpful |
| ☐ | ☐ | ☐ | ☐ | ☐ |

How precise did the interaction feel?

| Very imprecise | Imprecise | Neutral | Precise | Very Precise |
| ☐ | ☐ | ☐ | ☐ | ☐ |

Could you imagine a microscopist using this interface?

| Never | If they have to | Neutral | Sometimes | Everytime |
| ☐ | ☐ | ☐ | ☐ | ☐ |

Here you can write any comment you may have on the 3D Desktop interface.

```

```

The following questions concern the **VR** interface.

How close is the interface to the capabilities of state of the art microscope software?

| Worse | Slightly Worse | Comparable | Improvment | Big improvment |
| ☐ | ☐ | ☐ | ☐ | ☐ |

How enjoyable was your experience?

| Very Unenjoyable | Unenjoyable | Neutral | Enjoyable | Very Enjoyable |
| ☐ | ☐ | ☐ | ☐ | ☐ |

How helpful was the interface when performing the Many-Targets/Tube task?

| Very Unhelpful | Unhelpful | Neutral | Helpful | Very Helpful |
| ☐ | ☐ | ☐ | ☐ | ☐ |

How helpful was the interface when performing the Follow-Structure/Tree task?

| Very Unhelpful | Unhelpful | Neutral | Helpful | Very Helpful |
| ☐ | ☐ | ☐ | ☐ | ☐ |

How easy was it to orient yourself in the stage space?

| Very Difficult | Difficult | Neutral | Easy | Very Easy |
|---|---|---|---|---|
| ☐ | ☐ | ☐ | ☐ | ☐ |

How easy was it to move the focus to mark targets?

| Very Difficult | Difficult | Neutral | Easy | Very Easy |
|---|---|---|---|---|
| ☐ | ☐ | ☐ | ☐ | ☐ |

How natural and intuitive did it feel to use the desktop interface?

| Very unintuitive | Unintuitive | Neutral | Intuitive | Very intuitive |
|---|---|---|---|---|
| ☐ | ☐ | ☐ | ☐ | ☐ |

How difficult was it to learn how to use the interface?

| Very Difficult | Mildly Difficult | Neutral | Easy | Very Easy |
|---|---|---|---|---|
| ☐ | ☐ | ☐ | ☐ | ☐ |

Was the introduction helpful?

| Not at all | Unhelpful | Neutral | Helpful | Very helpful |
|---|---|---|---|---|
| ☐ | ☐ | ☐ | ☐ | ☐ |

How precise did the interaction feel?

| Very imprecise | Imprecise | Neutral | Precise | Very Precise |
|---|---|---|---|---|
| ☐ | ☐ | ☐ | ☐ | ☐ |

Could you imagine a microscopist using this interface?

| Never | If they have to | Neutral | Sometimes | Everytime |
|---|---|---|---|---|
| ☐ | ☐ | ☐ | ☐ | ☐ |

Here you can write any comment you may have on the VR interface.

## Tasks

Finally, a few questions about the tasks.

How often does a microscopist face the **Follow-Structure** task in their work?

| Never | Once | Sometimes | Monthly | Daily |
|---|---|---|---|---|
| ☐ | ☐ | ☐ | ☐ | ☐ |

How often does a microscopist face the **Many-Targets** task in their work?

| Never | Once | Sometimes | Monthly | Daily |
|---|---|---|---|---|
| ☐ | ☐ | ☐ | ☐ | ☐ |

Rank your favorite interface for solving the **Follow-Structure** task (1 = best, 4 = worst)

| 2D Desktop | 3D Desktop | VR | Other: | |
|---|---|---|---|---|
| | | | | |

Rank your favorite interface for solving the **Many-Targets** task? (1 = best, 4 = worst)

| 2D Desktop | 3D Desktop | VR | Other: | |
|---|---|---|---|---|
| | | | | |

## Comment

If you have anything you want to comment on please write it below.

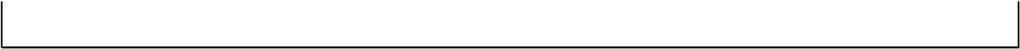

Thank you for participating. You may now choose a snack or stickers.



\end{document}